\documentclass[
    aps,
    prd,
    amsmath,
    amssymb,
    preprintnumbers,
    onecolumn,
    11pt,
    nofootinbib
]{revtex4}

\usepackage{mathtools}
\usepackage[
    a4paper,
    left=20mm,
    right=20mm,
    top=25mm,
    bottom=25mm
]{geometry}

\usepackage{amsfonts}
\usepackage{mathrsfs}
\usepackage{bm}

\usepackage{graphicx}

\usepackage{multirow}
\usepackage{diagbox}

\usepackage{xcolor}
\usepackage{comment}

\newcommand{\be}{\begin{equation}}
\newcommand{\ee}{\end{equation}}
\newcommand{\ba}{\begin{eqnarray}}
\newcommand{\ea}{\end{eqnarray}}

\renewcommand{\(}{\left(}
\renewcommand{\)}{\right)}

\newcommand\e{{\rm e}}

\newcommand{\Om}{\Omega}
\newcommand{\call}{{\cal L}}

\newcommand{\al}{\alpha}
\newcommand{\bt}{\beta}

\newcommand{\de}{\delta}

\newcommand{\bD}{{\bf D}}

\newcommand{\M}{{\cal M}}
\newcommand{\eff}{{\rm eff}}

\newcommand{\dph}{\dot{\phi}}
\newcommand{\ddph}{\ddot{\phi}}
\newcommand{\dddph}{\dddot{\phi}}

\newcommand{\dpi}{\dot{\pi}}

\newcommand{\dPh}{\dot{\Phi}}
\newcommand{\dPs}{\dot{\Psi}}
\newcommand{\dotH}{\dot{H}}

\newcommand{\lm}{\Lambda}
\begin{document}

\title{Spherical collapse and cluster number counts in DHOST theories that pass the constraints from gravitational waves}

\author{Sakdithut Jitpienka}
\email{sakdithutj62@nu.ac.th}
\affiliation{The Institute for Fundamental Study, Naresuan University, Phitsanulok 65000, Thailand}

\author{Khamphee Karwan}
\email{khampheek@nu.ac.th}
\affiliation{The Institute for Fundamental Study, Naresuan University, Phitsanulok 65000, Thailand}

\author{David F. Mota}
\email{d.f.mota@astro.uio.no}
\affiliation{Institute of Theoretical Astrophysics, University of Oslo, N-0315 Oslo, Norway}

\date{\today}

\begin{abstract}
We investigate the spherical collapse model and the abundance of galaxy clusters in a class of degenerate higher-order scalar--tensor (DHOST) theories in which gravitational waves do not decay into scalar perturbations and which are consistent with current constraints from gravitational-wave observations. We find that deviations from Einstein gravity can become significant at late times when the background universe is close to the scaling regime during the matter-dominated epoch. These deviations suppress the growth of linear matter perturbations on small scales while increasing the extrapolated linear density contrast at collapse, obtained from the spherical collapse model.
Using the analytic mass function, we compute the corresponding cluster number counts. The minimum mass threshold in the mass integration for each redshift bin is determined by matching the predicted number counts in the $\Lambda$CDM model with those inferred from the eROSITA survey. 
We find that the cluster abundance reaches its maximum at low redshift bin, and that the number of clusters in the highest redshift bin is suppressed as the deviation from Einstein gravity becomes larger.
The parameters of the theory are chosen such that the deviation from Einstein gravity at present is consistent with the local astrophysical bounds from  binary pulsar observations.
We find that even under such strict constraints, the upper bound on the deviation leads to lower predicted number counts compared with the Poisson error of the eROSITA survey results.
 However, this may be a consequence of the uncertainties in computing the number counts for the DHOST theories using the spherical collapse model and the analytical mass function.
{
 In general, it may be concluded that the suppression of the cluster number counts is a consequence of the enhancement of the extrapolated linear density contrast.
 }
\end{abstract}

\maketitle
\section{Introduction}

The origin of cosmic acceleration remains one of the central problems in modern cosmology. 
While the cosmological constant provides a simple phenomenological explanation consistent with current observations, 
modified gravity theories offer an alternative in which late-time acceleration arises from modifications of gravitational dynamics on large scales. 
Such scenarios can be tested through observations of large-scale structure, including the growth of matter perturbations and the abundance of galaxy clusters.

Degenerate higher-order scalar–tensor (DHOST) theories constitute a broad class of scalar–tensor theories that evade the Ostrogradsky instability through degeneracy conditions imposed on the Lagrangian. 
Recent multimessenger observations of gravitational waves, particularly GW170817 and its electromagnetic counterpart, have placed stringent constraints on the propagation speed of gravitational waves. 
These constraints significantly restrict the viable parameter space of DHOST theories and motivate detailed investigations of their cosmological implications.

One of the most powerful probes of modified gravity is the formation of nonlinear structures. 
In particular, the spherical collapse model provides a framework to connect the nonlinear evolution of matter overdensities with observable quantities such as the halo mass function and cluster number counts. 
Since these observables are sensitive to both the growth of perturbations and the underlying gravitational theory, they offer a robust way to test deviations from Einstein gravity.

In this work, we investigate the nonlinear evolution of matter perturbations and the resulting abundance of galaxy clusters in a class of quadratic DHOST theories consistent with gravitational-wave constraints.

This paper is organized as follows. 
In Sec.~II, we present the theoretical framework of DHOST theories and impose the constraints from gravitational-wave observations. 
In Sec.~III, we derive the background evolution equations and analyze the autonomous system. 
In Sec.~IV, we formulate the perturbation equations and study the growth of linear matter perturbations. 
In Sec.~V, we investigate the spherical collapse process. 
In Sec.~VI, we compute the cluster number counts and discuss the observational implications. 
Technical details and lengthy expressions are collected in the Appendix.

\section{DHOST theories}

Quadratic DHOST theories are described by the action \cite{Langlois:2015cwa}
\begin{align}
 S &= \int d^4x \sqrt{-g} \mathcal{L}_{\rm DHOST}\,,
\label{act1}
\end{align}
where $g$ is the determinant of the metric tensor $g_{\mu\nu}$.
The Lagrangian is given by
\begin{align}
 \mathcal{L}_{\rm DHOST} &\equiv 
 G_2(\phi,X) + G_3(\phi,X)\Box\phi 
 + G_4(\phi,X) R
    + \sum_{i=1}^5A_{i}(\phi,X)\mathcal{L}_{i}\,, 
    \label{eq:DHOST_action}
\end{align}
where $R$ is the Ricci scalar, and $G_2$, $G_3$, $G_4$, and $A_{i}$ are arbitrary functions of $\phi$ and 
$X \equiv - \nabla_\mu\phi\nabla^\mu\phi/2$.
The Lagrangians describing derivative couplings of the scalar field are
\begin{align}
 \mathcal{L}_1 \equiv \phi_{\mu\nu}\phi^{\mu\nu}, \quad
 \mathcal{L}_2 \equiv (\Box\phi)^2, \quad
 \mathcal{L}_3 \equiv (\Box\phi)\phi^\mu\phi_{\mu\nu}\phi^\nu,\quad
 \mathcal{L}_4 \equiv \phi^\mu\phi_{\mu\rho}\phi^{\rho\nu}\phi_\nu, \quad
 \mathcal{L}_5 \equiv (\phi^\mu\phi_{\mu\nu}\phi^\nu)^2\,,
\end{align}
where $\phi_\mu \equiv \nabla_\mu\phi$, $\phi_{\mu\nu} \equiv \nabla_\mu\nabla_\nu\phi$, and $\nabla_\mu$ is the covariant derivative compatible with $g_{\mu\nu}$.
In this work, we consider type-I DHOST theories, for which the degeneracy conditions impose the following three relations~\cite{DHOST2016:3}:
\begin{align}
A_2&=-A_1\,,
\label{ct2}\\
A_4&=\frac{1}{8\left(G_{4} - 2 A_2 X \right)^2}\Bigl[
- 32 X A_2^2+4\left( 3G_{4}+16XG_{4X}\right) A_2^2+\left( -32X^2G_{4X}+24XG_{4}\right) A_3A_2
\nonumber\\
&
- 4 X^2G_{4}A_3^2
-8G_{4X}\left( 3G_{4}+4XG_{4X}\right) A_2+8G_{4}\left( XG_{4X}-G_{4}\right)A_3+12G_{4}G_{4X}^2
\Bigr]\,,
\label{deg:a4}\\
A_5&=\frac{\left( -2G_{4X}+2A_2 -2 X A_3\right)\left( -2A_2^2-6XA_2A_3+2G_{4X}A_2+4G_{4}A_3\right)}{8\left(G_{4}-2XA_2\right)^2}
\,,\label{deg:a5}
\end{align}
where the subscripts $\phi$ and $X$ denote derivatives with respect to $\phi$ and $X$, respectively.
For quadratic DHOST theories, the propagation speed of gravitational waves (GWs) is given by \cite{deRham2016}
\be
c_T^2 = \frac{G_4}{G_4 - 2 X A_1}\,,
\label{Ct2}
\ee
where the speed of light is set to unity.
Observations from LIGO/VIRGO \cite{Monitor:2017mdv,TheLIGOScientific:2017qsa} constrain the deviation of the GW speed from unity to be
$ -3\times10^{-15} \le c_T - 1 \le + 7 \times10^{-16}$.
From Eq.~(\ref{Ct2}), this implies
\be
A_1 =0\,.
\label{gw:a1}
\ee
In addition, the decay of gravitational waves into scalar perturbations can be avoided by imposing \cite{Creminelli2018}
\be
A_3 =0\,.
\label{gw:a3}
\ee
Substituting Eqs.~(\ref{gw:a1}) and (\ref{gw:a3}) into Eqs.~(\ref{deg:a4}) and (\ref{deg:a5}), we obtain
\be
A_5 = 0\,,
\quad\mbox{and}\quad
A_4 = \frac{3 G_{4X}^2}{2G_4}\,.
\label{a4fin}
\ee
Hence, the action for quadratic DHOST theories that satisfy the gravitational-wave constraints can be written as \cite{Hirano:1903}
\be
S_G = \int d^4x \sqrt{-g} \left\{
G_2 + G_3\Box\phi
+ G_4 R 
+ \frac{ 3 G_{4X}^2}{2G_4} \phi^{\mu} \phi_{\mu \rho} \phi^{\rho \nu} \phi_{\nu}
\right\}\,.
\label{actGW2}
\ee
We define the total action for gravity and matter as
\be
S = S_G + S_m\,.
\label{actGW2-sm}
\ee
Here, $S_m$ is the action for minimally coupled matter.
In this work, we focus on models in which all functions depend only on $X$, and we adopt the following forms:
\be
G_2 = - c_1 X + c_2X^2\,,\qquad 
G_3 = - c_3 X\,,\qquad
G_4 = \frac{1}{2} + c_4 X^2\,,
\label{g-forms}
\ee
where the reduced Planck mass is set to $m_p = 1/ \sqrt{8\pi G} = 1$.

\section{Background evolution}

\subsection{Evolution equations}

For a spatially flat universe, we adopt the Friedmann–Lemaître–Robertson–Walker (FLRW) metric in the form
\be
ds^2  = -n^2 dt^2 + a^2 \delta_{ij} dx^i dx^j\,,
\ee
where the lapse function $n$ and the scale factor $a$ depend only on time.
Substituting this metric into the action (\ref{actGW2-sm}), we obtain
\ba
S &=& \int dt  \, n a^3 \left\{
-6 G_4 \left[ \frac{\dot {a}}{n a} 
+ \frac{G_{4 X}}{2 G_4}\frac{\dot\phi}{n^2} \frac{d}{dt}\left(\frac{\dot\phi}{n}\right) \right]^2  
-3 G_3 \frac{\dot {a}\dot\phi}{n^2 a}  
- G_3  \frac{1}{n} \frac{d}{dt}\left(\frac{\dot\phi}{n}\right)
+ G_2 \right\}
+ S_m\,,
\label{cosmodhost}
\ea
where a dot denotes derivative with respect to time.
Varying the action with respect to $n$ and $a$, and then setting $n = 1$, we obtain the background evolution equations, which can be written in the form 
\be
3 H^2 = \rho_\M + \rho_\eff\,,
\quad
2 \dotH + 3 H^2 = - P_\M + P_\eff\,,
\label{rhoeff}
\ee
where $ H \equiv \dot{a} / a $ is the Hubble parameter, $\rho_\M$ and $P_\M$ are the total energy density and pressure of matter and radiation in the universe, which are assumed to form a perfect fluid.
The effects of modification of gravity are encoded in the effective energy density $\rho_\eff$ and effective pressure $P_\eff$ of the scalar field,
where their expressions are given by Eqs.~\eqref{om-eff} and \eqref{w-eff}.
Variation of the action (\ref{cosmodhost}) with respect to $\phi$ yields an equation of motion of the form
\be
F(\ddddot{\phi}, \dddot\phi, \ddot\phi, \dot\phi, \ddot{H}, \dotH, H) = 0\,.
\label{kg}
\ee
Following the procedure described in Appendix~\ref{app:reduction}, this equation can be reduced to a quadratic equation for $\ddot{\phi}$:
\be
A \ddot\phi^2 + B \ddot\phi + C = 0\,,
\label{eq-ddp}
\ee
where the coefficients $A$, $B$, and $C$ are given by Eqs.~\eqref{defA} -- \eqref{defB}.

\subsection{Autonomous equations}

To analyze the background evolution, we substitute the forms of $G_2$, $G_3$, and $G_4$ given in Eq.~(\ref{g-forms}) into the evolution equations. 
The system can then be expressed in terms of the following dimensionless variables:
\begin{align} 
\label{dimless-param}
x_1^2 \equiv  \frac{2 c_1 X}{ H^2}\,,
\quad
x_2 \equiv  \frac{c_2 X^2}{ H^2}\,,
\quad
x_3 \equiv \frac{c_3 \dot\phi X}{ H}\,,
\quad
x_4 \equiv c_4 X^2\,,
\quad
\epsilon \equiv \frac{\ddot\phi}{H \dot\phi}\,,
\quad
\Omega_r \equiv \frac{\rho_r}{3 H^2}\,,
\quad
\Omega_m \equiv \frac{\rho_m}{3 H^2}\,. 
\end{align}
Here, $\rho_r$ and $\rho_m$ denote the energy densities of radiation and matter, respectively.
The autonomous system for these variables is given by
\begin{align}
x_{1 N} &= 2 x_{1} \left( \epsilon - h_N \right),
\label{x1n}\\
x_{2 N} &= 2 x_{2} \left( 2 \epsilon - h_N \right),
\label{x2n}\\
x_{3 N} &= x_{3} \left( 3 \epsilon - h_N \right),
\label{x3n}\\
x_{4 N} &= 4 x_{4} \epsilon, 
\label{x4n}\\
\Omega_{r N} &= -2 \Omega_{r}\left( 2 + h_N \right)\,,
\label{orn}\\
\Omega_{m N} &= - \Omega_{m}\left( 3 + 2 h_N \right)\,,
\label{omn}
\end{align}
where the subscript ${}_N$ denotes differentiation with respect to $N \equiv \ln a$,
and  $h_N \equiv \dot H / H^2$.
The quantity $\epsilon$ can be obtained by solving Eq.~(\ref{eq-ddp}) for $\ddot{\phi}$. 
We note that Eq.~(\ref{eq-ddp}) admits a single solution when $G_3 = 0$, while it yields two branches when $G_3 \neq 0$. 
In the latter case, $\epsilon$ takes the form
\be
\epsilon = a_1 \pm \sqrt{a_1^2 + a_2}\,,
\label{epsilon}
\ee
where the expressions for $a_1$ and $a_2$ are given in Eqs.~\eqref{defA1} and \eqref{defA2}.
In the case $x_3 = 0$, i.e., $G_3 = 0$, the expression for $\epsilon$ reduces to
\begin{align}
\epsilon = \frac{3 \left(2 x_4 - 1\right) \left(12 x_4 \Omega_m \left(-w_{M N} + 3 w_M^2 - w_M\right) + \left(6 x_4 - 1\right) x_1 + 4 x_2 \left(1 - 2 x_4\right)\right)}{-12 x_4 \left(2 x_4 + 3\right) \Omega_m \left(3 w_M - 1\right) + \left(-12 x_4^2 + 24 x_4 - 1\right) x_1 + 12 x_2 \left(1 - 2 x_4\right)^2}\,.
 \end{align}
For simplicity, we therefore set $G_3 = 0$ in the remainder of this work.
The effective equation-of-state parameter for the matter sector, $w_\M$, and its derivative are given by
\be
w_\M = \frac{1}{3} \frac{\Omega_r}{\Omega_r + \Omega_m}\,,
\quad
w_{\M N} = -\frac{\Omega_m \Omega_r}{3 \left(\Omega_m + \Omega_r\right)^2}\,. 
\ee
In terms of the dimensionless variables, we can write 
\be
h_N = h_N(x_1, x_2, x_4, \epsilon, \Omega_\M, w_\M)\,,
\label{dh2h2-dimen}
\ee
where the calculation used to obtain  this function is presented in Appendix \ref{app:reduction}.
Using Eqs.~(\ref{dimless-param}) and \eqref{om-eff},
Eqs.~(\ref{rhoeff}) can  be written as
\begin{align} 
\label{dens-param}
\Omega_\eff = 1 - \Omega_r - \Omega_m\,.
\end{align}

\subsection{Background evolution}

\begin{figure}
\centering
\includegraphics[width=0.6\textwidth]{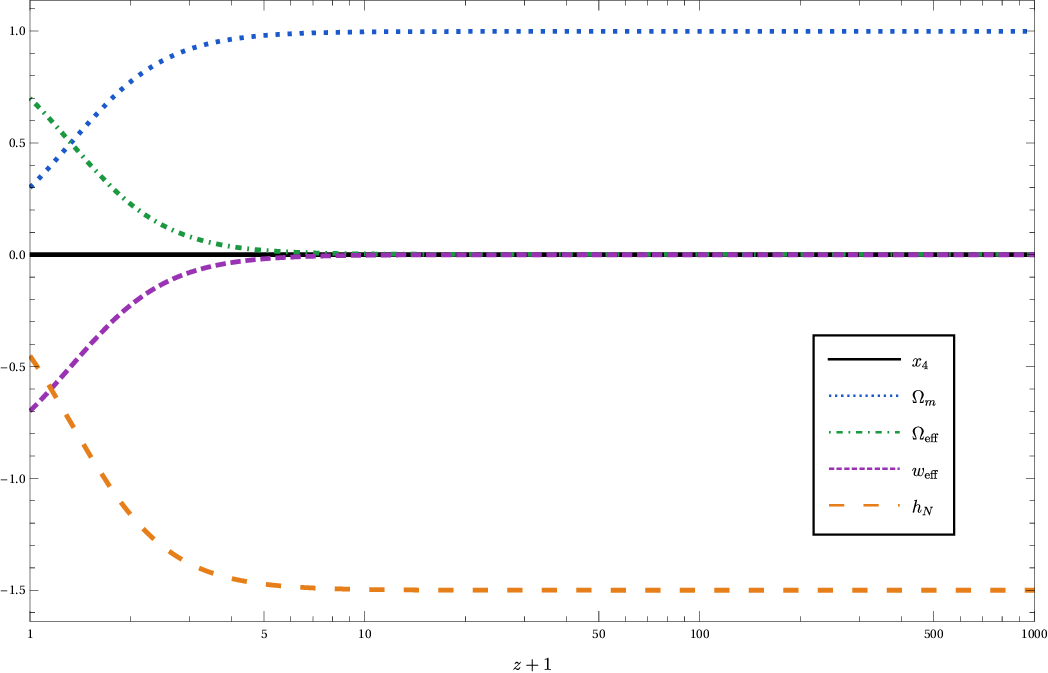}

\vspace{0.5cm} 

\includegraphics[width=0.6\textwidth]{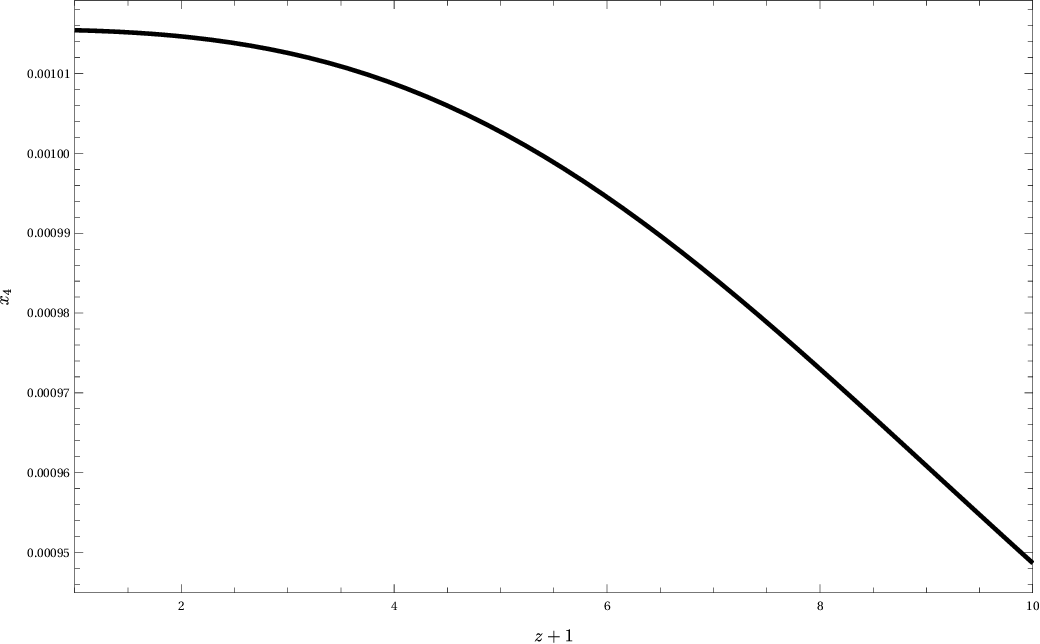}
\caption{
Evolution of $\Omega_m$, $x_4$, $\Omega_\eff$, $w_\eff$, and $h_N$ as functions of the redshift $z$, from the matter-dominated epoch to late times (top panel). The bottom panel zooms in on the late-time dynamical drift of the parameter $x_4$ covering the low-redshift regime up to $z \sim 9$ ($z+1 = 10$) using a linear scale.
\label{fig2}
}
\end{figure}

Since we are interested in the dynamics of spherical collapse, we neglect the contribution from radiation and denote the matter component by the subscript ${}_m$ in the following analysis. 
Deep in the matter-dominated epoch, we assume $\Omega_m \sim 1$, while $x_1$, $x_2$, and $x_4$ are much smaller than unity. 
Under these conditions, $\epsilon$, $h_N$, $\epsilon_N$, $\Omega_\eff$, and $w_\eff$ can be approximated as shown in Appendix~\ref{app:approx:dimless}.
The possible sequence of cosmic evolution can be understood by analyzing the fixed points of Eqs.~(\ref{x1n})--(\ref{x4n}). 
During the matter-dominated epoch, we have $\dotH/H^2 \simeq -3/2$, which leads to two fixed points. 
One corresponds to $x_1 = x_2 = x_4 = 0$, while the other is given by $x_1 = x_2 = \epsilon = 0$. 
The latter fixed point is of particular interest because $x_4$, which characterizes the deviation from Einstein gravity, does not necessarily vanish.
For this fixed point, Eqs.~(\ref{dens-param}) and (\ref{wphi}) give
\be
\Omega_m = 1 - 2 x_4,,
\quad
w_\eff = 0,.
\ee
Hence, this corresponds to a scaling solution.
In our analysis, we assume that the cosmic evolution starts from this scaling fixed point during matter domination. 
In the numerical integration, the system can evolve from the scaling solution toward a late-time accelerated phase if small deviations from the scaling fixed point are introduced, 
i.e., if $x_1$ and $x_2$ do not vanish exactly during the matter-dominated epoch.
Since the effects of modified gravity on the background dynamics are encoded in $\Omega_\eff$, $w_\eff$, and $x_4$, we use Eqs.~(\ref{omphi}) and (\ref{wphi}) to express $x_1$ and $x_2$ in terms of these quantities. 
We select the branch of solutions for $x_1$ and $x_2$ such that they remain positive in the range $-1 < w_\eff \leq 0$.
During the matter-dominated epoch, the decrease of $x_1$ and $x_2$ toward small values at a given redshift causes $\epsilon$ to approach zero; 
therefore, $x_4$ remains nearly constant according to Eq.~\eqref{x4n}.
when $\epsilon$ deviates from zero, $x_4$  decreases.
Hence, a sizable value of $x_4$ at late times can be achieved if the initial conditions are chosen such that $\epsilon \sim 0$. 
If the initial value of $\epsilon$ decreases toward zero, the scaling regime persists longer.
{The evolution of $\Omega_m$, $\Omega_\eff$, $w_\eff$, $h_N$, and $x_4$ is plotted in Fig.~(\ref{fig2}). 
The evolution shown in the plot is chosen such that $x_4 \sim 10^{-4}$ at present, which is consistent with the upper bound $x_4 \lesssim 10^{-4}$ from binary pulsar observations \cite{Hirano:1903}.}

\section{Evolution of the perturbations}
\subsection{Action for perturbations}

Since we are interested in the dynamics of spherical collapse, we focus on cosmological perturbations on small scales. 
The metric perturbations in the Newtonian gauge are given by
\begin{align}
ds^2 = -(1 + 2\Phi) dt^2 + a^2 (1 - 2\Psi) \delta_{ij} dx^i dx^j\,.
\label{ds2pert}
\end{align}
The scalar field and the matter energy density are decomposed into background and perturbed components as
\begin{align}
\label{decomposed}
\phi = \phi(t) + \pi(t, x)\,,\quad
\rho_m = \bar{\rho}_m(t) + \delta\rho_m(t, x)\,.
\end{align}
Using these decompositions, the action (\ref{actGW2-sm}) can be expanded as
\be
\delta S = \int d^4 x \left( \call_2 + \call_3 + \call_4 \right)\,,
\label{delS}
\ee
where the subscripts $2$, $3$, and $4$ denote the perturbative order. 
The explicit expressions for these perturbed Lagrangians are given in Eqs.~\eqref{lp2} -- \eqref{lp4}.
Under the quasi-static approximation on small scales \cite{Kimura:2011dc, Koyama:2013paa}, we retain the dominant contributions as follows. 
Since the perturbation variables $\Phi$, $\Psi$, and $\pi$ are small, nonlinear terms in these fields, as well as terms involving first-order derivatives of $\pi$, can be neglected. 
However, terms containing second or higher-order spatial derivatives can become large on small scales and must therefore be retained.
Furthermore, time derivatives of the perturbations are negligible compared to spatial derivatives. 
For example, terms of the form $\dot{\chi}\,\partial^2 \chi$ can be neglected relative to $\partial_i \chi \, \partial^i \chi$, where $\chi$ denotes any perturbation variable.
Varying the perturbed action (\ref{delS}) with respect to $\Phi$, $\Psi$, and $\pi$, we obtain the equations of motion for the perturbations, which are presented in Appendix~\ref{app:perteom:all}.

\subsection{Growth of linear matter perturbations on small scales}

\begin{figure}
\includegraphics[width=0.6\textwidth]{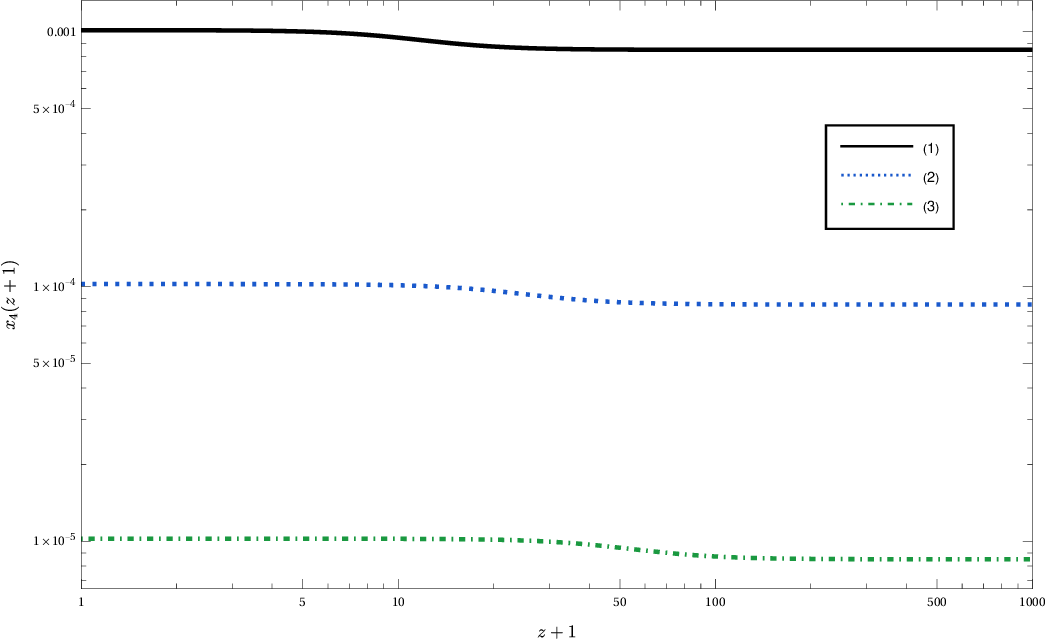} 
\qquad
\includegraphics[width=0.6\textwidth]{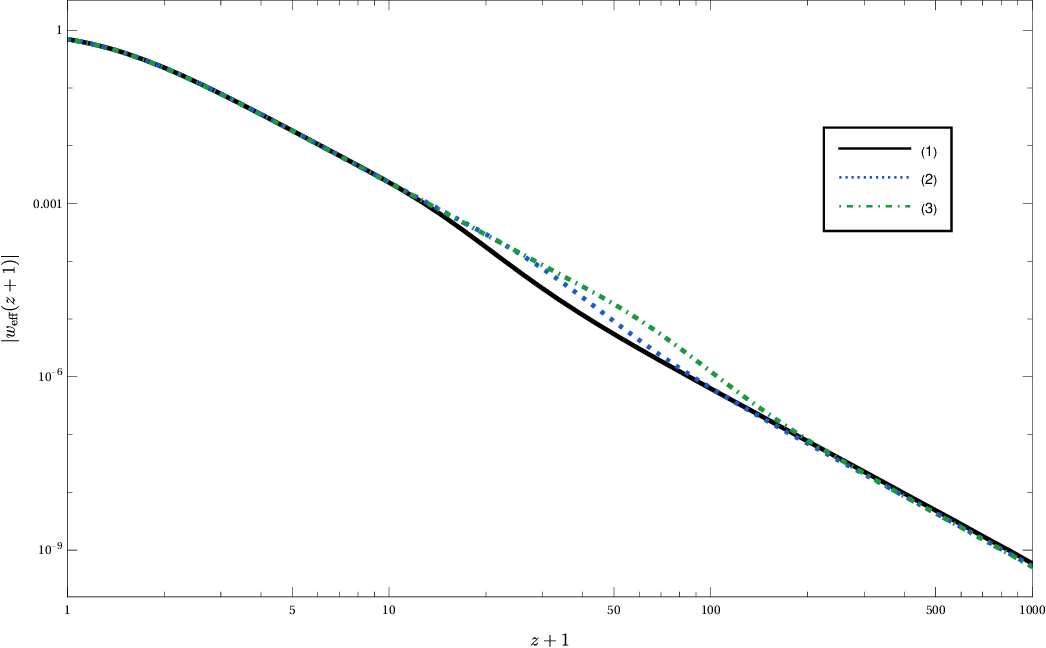}
\caption{
{Evolution of $x_4$ and $|w_\eff|$. 
Superscripts ${}^{(1)}$, ${}^{(2)}$, and ${}^{(3)}$ represent the cases where $x_{4,0} = 10^{-3}, 10^{-4}$, and $10^{-5}$, respectively. 
\label{fig:34}}
}
\end{figure}

At linear order in perturbations, Eqs.~(\ref{eomphi})--(\ref{eomfield}) reduce to a closed system. 
Following the method outlined in Appendix~\ref{app:linpert}, we obtain, on small scales,
\be
\Psi_{, i}^{, i} = C_1 \delta_m + C_2 \dot{\delta}_m + C_3 \ddot{\delta}_m\,. 
\label{psisolF}
\ee
On small scales, conservation of the energy-momentum tensor of matter leads to
\be
\ddot{\delta}_m + 2 H \dot{\delta}_m = \Psi^{, i}_{, i}\,, 
\label{ddotdel0}
\ee
where $\delta_m \equiv \delta\rho_m / \bar{\rho}_m$ is the matter density contrast.
Substituting Eq.~(\ref{psisolF}) into Eq.~(\ref{ddotdel0}), we obtain
\be
\left(1 - \frac{C_3}{H^2 a^2}\right) \delta_{m\, NN} 
+ \left(2 + \frac{\dotH}{H^2} - \frac{C_2}{H^2 a^2}\right)\delta_{m\, N} 
- \frac{C_1}{H^2 a^2}\delta_m = 0\,. 
\label{ddotdelEq}
\ee
The coefficients in Eq.~(\ref{ddotdelEq}) can be expressed in terms of the dimensionless variables $x_1$, $x_2$, $x_4$, and $\Omega_m$.
The evolution equation~(\ref{ddotdelEq}) is solved numerically with initial conditions set during the matter-dominated era. 
Under the approximation $x_1, x_2 \ll x_4$, Eq.~(\ref{ddotdelEq}) reduces to
\begin{align}
\delta_{m\, NN} + \left(\frac{1}{2} + 6 \Omega_m x_4\right)\delta_{m\, N} 
+ \left(3 \Omega_m (2 + \Omega_m) x_4 - \frac{3}{2} \Omega_m\right)\delta_m = 0\,.
\end{align}
Assuming that $\Omega_m$ and $x_4$ are approximately constant during the matter-dominated epoch due to the scaling behavior, the solution takes the form
\begin{align}
\delta_m =  A_+ \e^{\lambda_+ N} + A_- \e^{\lambda_- N}\,, 
\label{delm-init}
\end{align}
where $A_\pm$ and $\lambda_\pm$ are constants, and
\be
\lambda_{\pm} = \frac{1}{4} \left(-1 - 12 \Omega_m x_4 \pm \sqrt{1 + 24 \Omega_m - 72 \Omega_m x_4 - 48 \Omega_m^2 x_4 + 144 \Omega_m^2 x_4^2}\right)\,,
\label{lambda}
\ee
where the growing solution corresponds to $\lambda_+$.
In the numerical integration, we set $\delta_m = 1$ and $\delta_{m N} = \lambda_+$ at the initial time. 
To ensure that $x_4$ has a sizable value at late time, its value is chosen according to the scaling solution, $x_4 \simeq (1 - \Omega_m)/2$. 
{For comparison, we consider three DHOST background cosmologies, characterized by the present-day values $x_{4,0}=10^{-3}, 10^{-4},$ and $10^{-5}$. 
Since $w_\eff \sim 0$ during the matter-dominated epoch, 
$x_4$ remains nearly constant for all cases from the matter-dominated epoch onward for an extended period, as shown in Fig.~(\ref{fig:34}).} 
{The evolution of $\Omega_m$ differs from that in the $\Lambda$CDM model by less than one percent throughout cosmic history. 
This is because $w_\eff$ drops from zero to $-1$ while $\Omega_m$ is still non-negligible.
 Consequently, the background evolution, characterized by $\dot{H}/H^2$, is nearly identical for all choices of initial conditions.}

\begin{figure}
\includegraphics[width=0.6\textwidth]{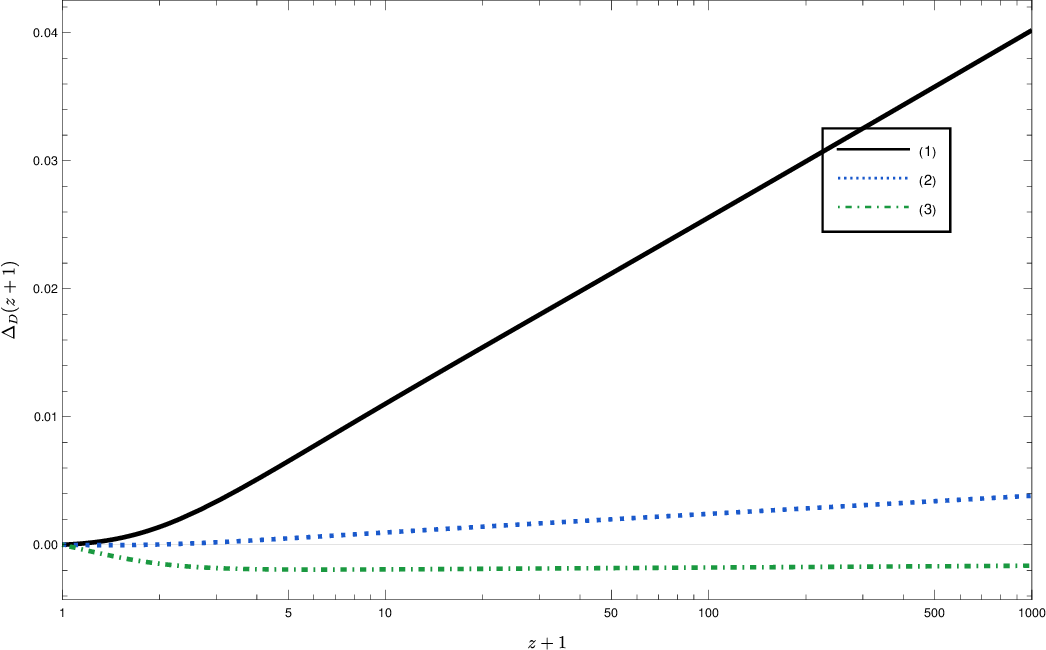}
\caption{
Relative difference in the growth factor, $\Delta_D$, as a function of redshift. 
The labels ${(1)}$--${(3)}$ correspond to the same present value of $x_4$ as in Fig.~\ref{fig:34}.
\label{figcontrast}
}
\end{figure}

In Fig.~\ref{figcontrast}, we show the relative difference in the growth factor as a function of redshift,
\be
\Delta_D(z) \equiv \frac{D(z) - D^{(\Lambda)}(z)}{D^{(\Lambda)}(z)}\,,
\ee
where quantities with the superscript ${}^{\Lambda}$ correspond to the $\Lambda$CDM model with the same cosmological parameters as those considered in the DHOST theory.
The growth factor is defined as
\be
D(z) \equiv \frac{\delta_m(z)}{\delta_m(z = 0)}\,.
\label{def:growth}
\ee
The results indicate that the growth rate of the matter density contrast, $\delta_m$, is suppressed in DHOST theories as $x_4$ increases. This behavior can be understood from Eq.~(\ref{lambda}), where a larger $x_4$ reduces $\lambda_+$. 
Since $x_4$ quantifies the deviation from Einstein gravity, this implies that modified gravity suppresses the growth of matter perturbations in the class of DHOST theories considered here.

\section{Spherical collapse}

To investigate spherical collapse, 
the evolution equation for the density contrast up to second order can be derived from the conservation equation $\nabla_\mu T^\mu_\nu = 0$. 
In this calculation, we assume a top-hat density profile of the form
\be
\rho_m(t, r) =
\left\{
\begin{array}{cc}
\bar{\rho}_m (1 + \delta_m(t)) & r \leq R \\
\bar{\rho}_m & r > R
\end{array}
\right.\,,
\ee
where $\bar{\rho}_m$ is the background matter energy density, and $\delta_m(t) \equiv (\rho_m - \bar{\rho}_m)/\bar{\rho}_m$ represents the overdensity within a spherical region of radius $R$.
{
Due to the existence of the fifth force and screening mechanisms in DHOST theories, 
the top-hat assumption may not be applicable to a realistic collapse process. 
This is because the efficiency of the screening mechanism depends on the mass and radius of the collapsing region. 
Hence, the screening mechanism may become effective only during the late stage of the collapse. 
The transition from the unscreened to the screened regime can lead to the breakdown of the top-hat density profile. 
A more detailed discussion is presented in the next section and in Appendix~\ref{app:collapse}.
}
Under the top-hat assumption, the second-order evolution equation for the density contrast is given by~\cite{Schmidt:09}
\be
\ddot{\delta}_m + 2 H \dot{\delta}_m 
- \frac{4}{3}\frac{\dot{\delta}_m^2}{1+\delta_m}
= (1+\delta_m) \Psi^{, i}_{, i}\,.
\label{eq:collapseeqn}
\ee
The quantity $\Psi^{, i}_{, i}$ up to second order is obtained from Eqs.~\eqref{eomphi}--\eqref{eomfield}. 
Following the analysis of Ref.~\cite{Hirano:1903}, we compute the nonlinear contributions to $\nabla^2 \Psi$ under the assumption of spherical symmetry. 
Spatial derivatives are then expressed in terms of derivatives with respect to the comoving radial coordinate $r$, so that Eq.~\eqref{eq:collapseeqn} becomes
\be
\delta_{m\,NN} 
+ \left(2 + \frac{\dotH}{H^2}\right)\delta_{m\,N} 
- \frac{4}{3}\frac{\delta_{m\,N}^2}{1+\delta_m}
= \frac{1}{H^2} (1+\delta_m)\,\nabla^2 \Psi(\delta_{m\,NN}, \delta_{m\,N}, \delta_m)\,.
\label{eq:collapseeqn1}
\ee
The derivation of Eq.~\eqref{eq:collapseeqn1} is presented in Appendix~\ref{defA}. 
We solve Eq.~\eqref{eq:collapseeqn1} numerically by specifying initial conditions during the matter-dominated epoch.
During initial period, the nonlinear terms in Eq.~\eqref{eq:collapseeqn1} can be neglected, and the equation reduces to Eq.~\eqref{ddotdelEq}. 
Therefore, the initial condition in Eq.~\eqref{delm-init} can be adopted.

The numerical solution of Eq.~\eqref{eq:collapseeqn1} allows us to determine the extrapolated linear density contrast $\delta_c$, which is a key parameter in the halo mass function. 
This quantity is obtained as follows. 
We first solve Eq.~\eqref{eq:collapseeqn1} with initial conditions set at high redshift such that $\delta_m$ initially follows the linear solution. 
We then tune the initial value of $\delta_m$ such that $\delta_m \rightarrow \infty$ (i.e., collapse) at a given redshift $z$. 
Using the same initial conditions in the linear equation~\eqref{ddotdelEq}, we compute the corresponding linear density contrast at the collapse redshift, which defines $\delta_c$.

\begin{figure}
\includegraphics[width=0.6\textwidth]{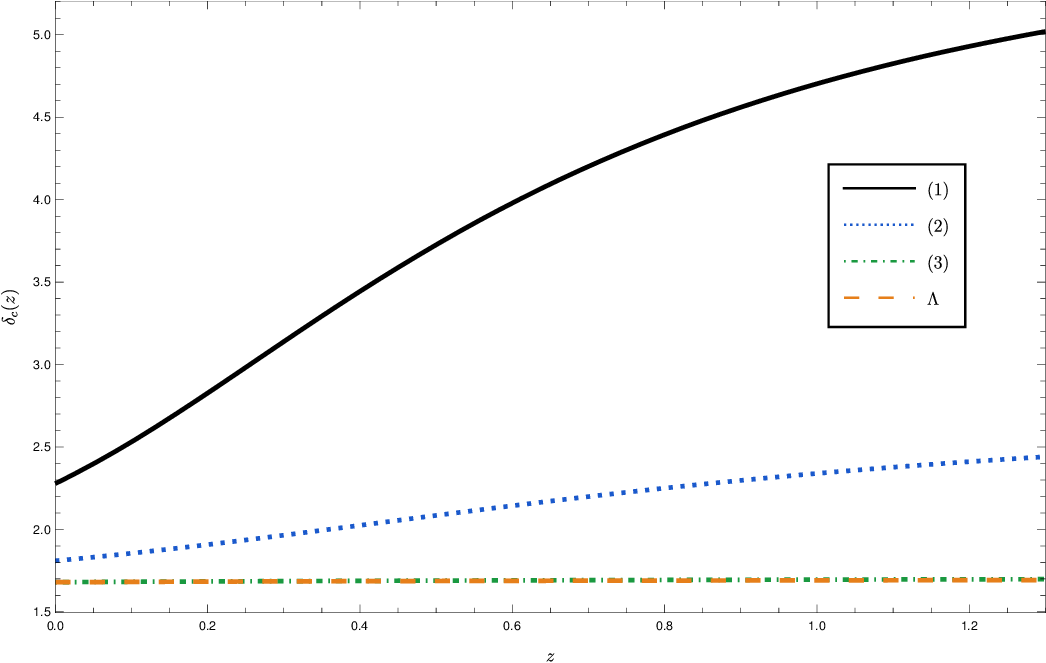}
\caption{
{Evolution of the extrapolated linear density contrast $\delta_c(z)$ for different DHOST background cosmologies as a function of redshift. 
The labels ${(1)}$--${(3)}$ correspond to the different DHOST background cosmologies as defined in Fig.~\ref{fig:34}, while ${\Lambda}$ denotes the $\Lambda$CDM model.
\label{figdelc}
}}
\end{figure}

From Fig.~\ref{figdelc}, we find that at a given redshift, $\delta_c$ increases as $x_4$ increases. 
Different from  the behavior of the growth rate, the extrapolated linear density contrast is enhanced by deviations from Einstein gravity. 
For cases ${(3)}$, $\delta_c$ is nearly identical to that of the $\Lambda$CDM model due to the smallness of $x_4$.

\section{Cluster number counts}
\label{sec6}

The abundance of collapsed structures can be estimated using the halo mass function, incorporating the linearly extrapolated density contrast obtained from the spherical collapse model. 
Within this framework, the comoving number density of objects with masses in the interval $[M, M+dM]$ is given by
\be
n(M)\, dM  =
- \frac{\tilde{\rho}_m}{M^2}\frac{d \ln\sigma}{d \ln M} f(\sigma),
\label{diffrel}
\ee
where $\tilde{\rho}_m \equiv a^3 \rho_m$ is the comoving matter energy density.
The root-mean-square (RMS) linear overdensity $\sigma$ for a region of radius $R$ enclosing a mass $M$ is approximated as~\cite{Viana:1999}
\be
\sigma(R,z)=\sigma_8\left(\frac{R}{8 h^{-1} {\rm Mpc}}\right)^{-\gamma(R)} D(z)\,.
\label{varinf}
\ee
The function $\gamma(R)$ is given by
\be
\gamma(R)= (0.3 \Omega_m h + 0.2)
\left[2.92+\log_{10}\left(\frac{R}{8 h^{-1} {\rm Mpc}}\right)\right]\,.
\ee
The mass function $f(\sigma)$ is defined as the fraction of mass in collapsed regions per unit redshift interval.
{
Since no mass functions specifically developed for DHOST theories are currently available, 
we investigate the effects of deviations from Einstein gravity on cluster number counts using the available analytical mass functions. 
In our analysis, we consider both the Press--Schechter (PS) and Sheth--Tormen (ST) mass functions.
}
The analytical mass function proposed by Press--Schechter (PS) \cite{Press:1974} takes the form
\begin{equation}
\label{ps_massfn}
f_{\rm PS}(\sigma) = \sqrt\frac{2}{\pi} \frac{\delta_{\rm crit}}{\sigma} e^{-\frac{\delta_{\rm crit}^2}{2\sigma^2}}\,.
\end{equation}
The improvement of the Press--Schechter model, the Sheth--Tormen (ST) mass function, proves more compatible with the N-body simulations used to study the abundance of dark matter halos in the $\Lambda$CDM model. 
This mass function, which was proposed in \cite{Sheth:1999}, is given by
\begin{equation}\label{stfunc}
f_{\rm ST}(\sigma) = A\sqrt{\frac{2a}{\pi}}
\left[1+\left(\frac{\sigma^2}{a\delta_c^2}\right)^p\right]
\frac{\delta_c}{\sigma}
\exp\left[-\frac{a\delta_c^2}{2\sigma^2}\right]\,.
\end{equation}
Here, $\delta_c$ is the linearly extrapolated density contrast at collapse. 
The parameters are set to $A=0.3222$, $a=0.707$, and $p=0.3$ \cite{Reed:2006rw}. 

\begin{figure}[htbp]
\centering
\includegraphics[width=0.45\textwidth]{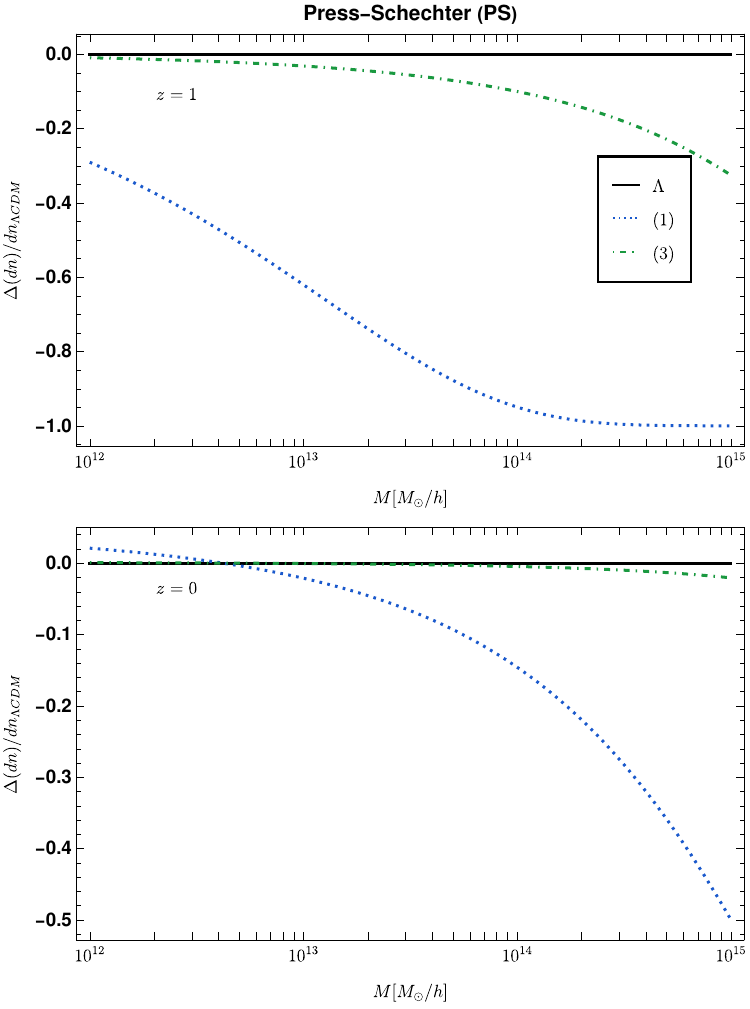}
\hfill
\includegraphics[width=0.45\textwidth]{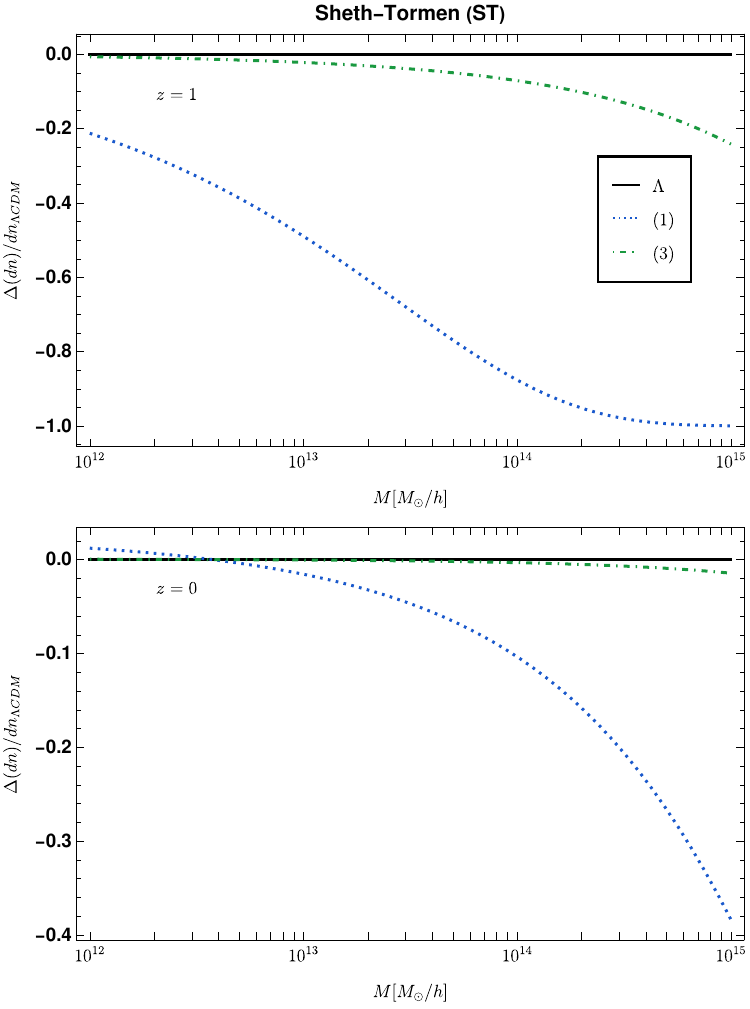}
\caption{The comoving number density of galaxy clusters as a function of mass for the Press--Schechter (left) and Sheth--Tormen (right) mass functions. The results indicate a suppression of cluster abundance in DHOST theories relative to the $\Lambda$CDM model.
}
\label{fig:nm}
\end{figure}

{
It is well known that the ST mass function provides a better fit to the results of $\Lambda$CDM $N$-body simulations than the PS mass function given by Eq.~(\ref{ps_massfn}). 
This improvement arises primarily because the Sheth--Tormen formalism is motivated by ellipsoidal collapse, 
which provides a more realistic description of the intrinsically non-spherical, or triaxial, nature of gravitational collapse. 

In this work, our primary objective is to investigate the qualitative impact of modified gravity on the abundance of collapsed objects rather than to provide a precision prediction for the halo mass function. 
For this purpose, the spherical-collapse model provides a sufficiently simple and transparent framework for isolating the main effects of the modified gravitational dynamics. 
Nevertheless, replacing the PS mass function with the standard ST prescription in DHOST theories does not necessarily provide a straightforward improvement. 
In fact, several additional complications arise when extending the Sheth--Tormen formalism, which was calibrated for $\Lambda$CDM, to modified gravity. 
Here, we highlight two important issues.

First, the ST mass function contains three phenomenological parameters, $A$, $a$ and $p$, whose values are determined by fitting to $N$-body simulations of $\Lambda$CDM cosmology. 
These parameters effectively encode the departure of the collapse dynamics from the idealized spherical-collapse picture. 
Although ellipsoidal collapse provides a more realistic description of halo formation, the magnitude and nature of the departure from spherical symmetry can depend on the underlying theory of gravity. 
In modified gravity, the evolution of overdense regions, the collapse threshold, and the relative importance of the fifth force can all differ from their $\Lambda$CDM counterparts. 
Consequently, there is no a priori reason to expect the $\Lambda$CDM-calibrated values of $A$, $a$ and $p$ to remain valid in DHOST theories. 
Ideally, these parameters should be recalibrated using dedicated $N$-body simulations performed in the corresponding modified-gravity model. 
Without such a recalibration, applying the standard Sheth--Tormen parameters may introduce additional uncertainties that are difficult to disentangle from the genuine effects of modified gravity.

Second, the hierarchical nature of structure formation raises a more fundamental issue for applying the top-hat collapse picture to screened modified gravity. 
In the standard hierarchical scenario, massive halos are assembled through the continuous merging and accretion of smaller, denser structures. 
Therefore, the assumption that an overdense region can be treated as a single top-hat perturbation throughout the entire collapse history may not provide an adequate description of the formation of a realistic halo, 
particularly when screening mechanisms are important. 
This issue is especially relevant because the efficiency of screening generally depends on the local density, gravitational potential, or characteristic size of the collapsing structures.

To illustrate this point, consider the formation of a very massive halo. 
In the idealized spherical top-hat collapse model, the density of the collapsing region initially remains relatively low, 
and the screening mechanism becomes significant only at relatively late stages of the collapse, when the density has increased sufficiently. 
The resulting modification of the collapse dynamics is therefore concentrated toward the final stages of halo formation. 
In a realistic hierarchical formation scenario, however, the same massive halo is assembled from smaller progenitors through a sequence of mergers and accretion events. 
These progenitor halos can reach sufficiently high densities, or possess sufficiently deep gravitational potentials, for the screening mechanism to become effective at much earlier times. 
Their subsequent mergers and accretion therefore carry information about the screened dynamics into the formation history of the final massive halo. 
As a consequence, screening can influence the assembly of a massive halo well before the density of the final, idealized top-hat region becomes high enough for screening to operate.

These considerations do not invalidate the use of the spherical-collapse and analytic mass function.
Rather, they emphasize that the resulting mass functions should be interpreted as qualitative estimates of the impact of modified gravity on halo abundance. 
A quantitative prediction would require a self-consistent treatment of ellipsoidal collapse, a recalibration of the Sheth--Tormen parameters in the modified-gravity model, and a proper account of the hierarchical assembly of halos and the associated evolution of the screening mechanism. 
We leave such a detailed investigation to future work.
}
{
In Fig.~(\ref{fig:nm}), we plot the number density of galaxy clusters using the PS and ST mass functions. 
The plots show that the number density of clusters is suppressed in DHOST theories compared with the $\Lambda$CDM model. 
The main contribution to this suppression comes from the enhancement of the extrapolated linear density contrast in DHOST theories, 
because this quantity appears in the exponential factor of the mass functions. 
For a given value of the extrapolated linear density contrast, the suppression is larger for more massive clusters because $\sigma(R,z)$ decreases as the mass enclosed within the radius $R$ increases. 
Comparing the cluster number densities at different redshifts, 
the plots show that the suppression becomes stronger at higher redshifts.
The above features of the cluster number density are common to both the PS and ST mass functions.
}

The number of galaxy clusters in a redshift bin $\Delta z$ with mass larger than $M_{\rm min}$ is obtained by integrating Eq.~\eqref{diffrel} over mass:
\be
N = f_{\rm sky} \frac{dV_e}{dz}\, \Delta z
\int_{M_{\rm min}}^\infty n(M)\, dM\,,
\label{dNdz}
\ee
where $f_{\rm sky}$ is the observed sky fraction, 
and $dV_e/dz \equiv 4\pi r(z)^2/H(z)$ is the comoving volume element per unit redshift, with $r(z)$ being the comoving distance. 
The volume element is evaluated at the midpoint of each redshift bin.

For the cluster number count analysis, we adopt a sky coverage of $f_{\rm sky} = 0.31$ (corresponding to $\approx 12{,}791~{\rm deg}^2$), consistent with the eROSITA survey~\cite{Bulbul:2024}. 
The mass threshold $M_{\rm min}$ in each redshift bin is chosen such that the predicted cluster counts in the $\Lambda$CDM model reproduce the eROSITA results.

\begin{figure}
\includegraphics[ width=0.6\textwidth, angle=0]{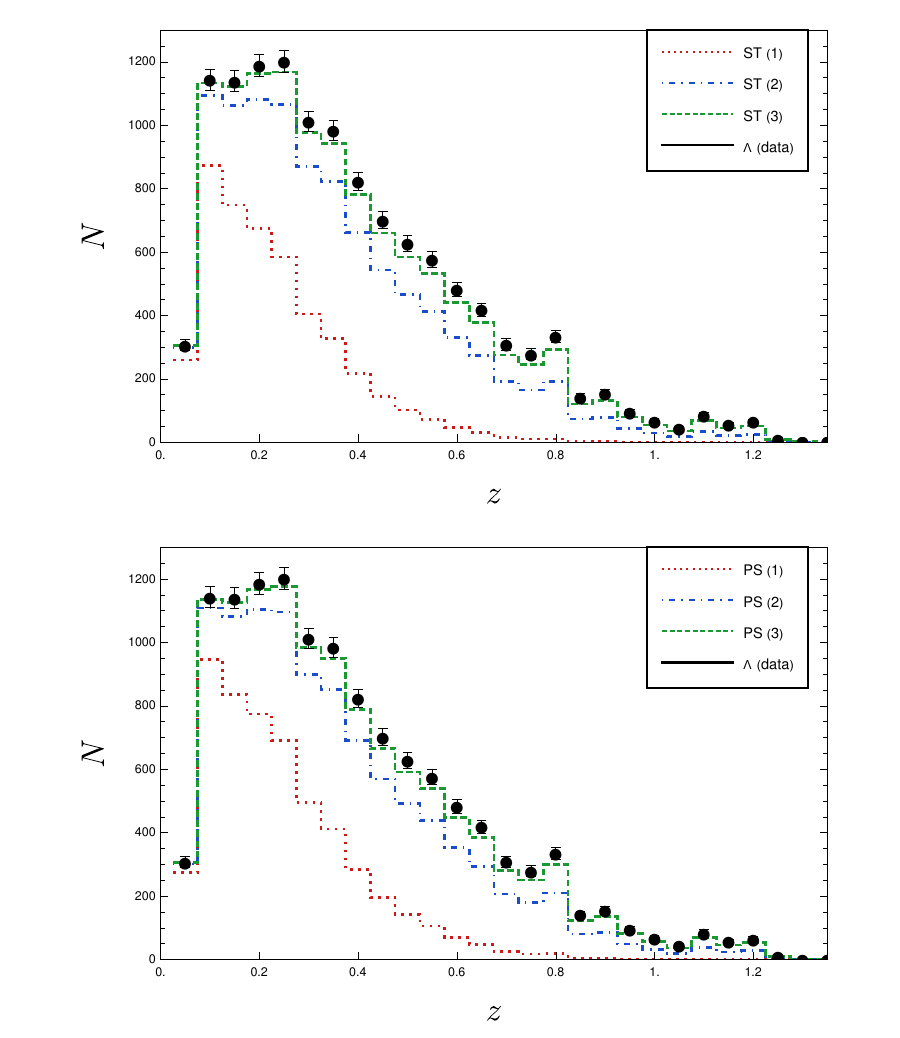}
\caption{
\label{dNdzABC}
Number of galaxy clusters $N$ per redshift bin $\Delta z$.
The upper panel shows the results obtained using the Sheth--Tormen (ST) mass function, while the lower panel corresponds to the Press--Schechter (PS) mass function.
{The labels ${(1)}$--${(3)}$ correspond to the different DHOST background cosmologies as defined in Fig.~\ref{fig:34}, while ${\Lambda}$ denotes the $\Lambda$CDM model.}
}
\end{figure}

From Fig.~\ref{dNdzABC}, we find that the number of galaxy clusters per redshift bin in DHOST models is significantly suppressed relative to the $\Lambda$CDM model, while the peak of the redshift distribution shifts toward lower redshifts for both choices of the mass functions.
{
The suppression of cluster abundances in DHOST theories is primarily driven by modifications of the collapse threshold $\delta_c$ and the growth factor $D(z)$, both of which affect the halo mass function.
In particular, a larger value of $\delta_c$, together with a suppressed growth rate, enhances the ratio $\delta_c^2 / \sigma^2$ in the exponential term of the mass function.
As a result,  the abundance of massive halos is reduced and the formation of nonlinear structures is delayed relative to the $\Lambda$CDM case.
We may conclude that the cluster number counts in the DHOST theory considered here is suppressed relative to the $\Lambda$CDM model independent of the choice of analytic mass function. 
This is because the exponential factor representing the Gaussianity of the matter density field under the assumptions of the Press-Schechter formalism should generally appear in the mass functions. 
The error bars shown for the $\Lambda$CDM model correspond to Poisson uncertainties of the eROSITA survey results,
which are estimated as $\sigma_N=\sqrt{N}$ for each redshift bin, where $N$ is the number of clusters in each redshift bin.
The predicted cluster abundances for cases ${(1)}$ and ${(2)}$ lie outside the Poisson error bars over the entire redshift range considered, whereas part of the peak region for case ${(3)}$ remains consistent with the corresponding error bars.
}
However, it cannot be concluded that the cluster number count observations prefer low $x_4$ compared with the bounds from binary pulsar observations,
because the preference for low $x_4$ could be a consequence of uncertainties in computing the number counts for DHOST theories using the spherical collapse model and the analytical mass functions.
Moreover, even though the $\Lambda$CDM model can emulate the number of clusters in each redshift bin from the eROSITA results, the Poisson uncertainties in the $\Lambda$CDM model do not reflect the actual observational errors.


\section{Conclusions}

In this work, we have investigated the nonlinear evolution of matter perturbations and cluster number counts within a class of quadratic DHOST theories that satisfy current gravitational-wave constraints. By applying the results from the spherical collapse model to the analytic mass function, we investigate the effects of DHOST theories on nonlinear structure formation.

Our findings reveal that deviations from Einstein gravity, quantified by the parameter $x_4$, suppress the growth rate of linear matter perturbations while enhancing the extrapolated linear density contrast at collapse. 
The increase in the extrapolated linear density contrast implies that larger initial overdensities are required for gravitational collapse to occur in DHOST cosmologies compared to the $\Lambda$CDM model. These two physical effects---the slowed growth of perturbations and the elevated collapse threshold---synergistically reduce the predicted abundance of galaxy clusters. 
The suppression is most pronounced as $x_4$ increases, with the peak of the cluster redshift distribution decreasing in amplitude and shifting toward lower redshifts, 
signaling a delayed formation of massive structures.

{ Setting the model parameters such that the present-day value $x_{4,0}$ is in agreement with the constraints from binary pulsar observations,}
the upper bound on $x_4$ from observations leads to lower predicted number counts compared with the $\Lambda$CDM model, emulating the eROSITA survey results.
However, this does not imply that the observed cluster number counts favor lower values of $x_4$ compared to other local astrophysical bound because of uncertainties in computing the number counts for the DHOST theories using the spherical collapse model and the analytic mass function. 
Although theoretical uncertainties regarding fitting mass functions remain, this work provides a necessary step toward understanding nonlinear structure formation in DHOST cosmologies consistent with gravitational-wave constraints. 
We hope that future developments in nonlinear simulations and large-scale structure surveys will further clarify the observational signatures of these theories beyond linear perturbations.

\section*{Acknowledgments}

We thank Ø. Elgarøy for useful comments on the manuscript. DFM acknowledges support from the Research Council of Norway and the resources provided by UNINETT Sigma2, the National Infrastructure for High-Performance Computing and Data Storage in Norway. 
This work was partially supported by Reinventing University Program 2026, The Office of the Permanent Secretary of the Ministry of Higher Education, Science, Research and Innovation (MHESI), and Naresuan University, Thailand 
[Grant No. R2569A013]

\appendix

\section{Background equations}
\label{bg:part}

Varying the action in Eq.~\eqref{cosmodhost} with respect to $n$ and $a$, and then setting $n = 1$,
we get the following evolution equations
\ba
&& -\frac{3 X}{G_4} \left(2 \dot{\phi } G_{4 X}^2 \dddot\phi +8 G_4 H^2 G_{4 X}+6 H \dot{\phi } G_{4 X}^2 \ddot{\phi }-2 G_4 H \dot{\phi } G_{3 X}+4 G_4 \dotH G_{4 X}+G_{4 X} \ddot{\phi }^2 \left(2 \dot{\phi }^2 G_{4 X X}+G_{4 X}\right)\right) \nonumber \\
&& +6 H \left(G_4 H+\dot{\phi } G_{4 X} \ddot{\phi }\right)+\frac{6 X^2 G_{4 X}^3 \ddot{\phi }^2}{G_4^2}-2 X G_{2 X}+G_2-\rho _m
= \rho_\M\,,
\label{e00}\\
&& 2 \dot{\phi } G_{4 X} \dddot\phi +6 G_4 H^2+4 H \dot{\phi } G_{4 X} \ddot{\phi }+4 G_4 \dotH+2 G_{4 X} \ddot{\phi }^2+\frac{X \ddot{\phi }}{G_4} \left(-3 G_{4 X}^2 \ddot{\phi }+4 G_4 G_{4 X X} \ddot{\phi }+2 G_4 G_{3 X}\right) \nonumber\\
&& +G_2+\rho _m w_m
= - P_\M\,.
\label{eii}
\ea
The above equations can be written in the form of the Friedmann and acceleration equations for the Einstein gravity as shown in Eq.~\eqref{rhoeff} 
by defining the effective energy density $\rho_\eff$ and effective pressure $P_\eff$ of the scalar field which can be expressed in terms of the dimensionless variables as 
\begin{align}
\Omega_\eff & \equiv \frac{\rho_\eff}{3 H^2} \nonumber \\
&= \frac{2 x_4}
{\left(2 x_4+1\right)^2} \left(4 x_4^2 \left((8 \epsilon +4) h_N+8 \epsilon_N + 12 \epsilon^2 + 20 \epsilon + 7\right)+4 x_4 \left(4 (\epsilon +1) h_N + 4 \epsilon_N + 14 \epsilon^2+8 \epsilon +7\right) \right.
\nonumber\\
&\left.
+4 h_N-4 \epsilon + 7\right) 
- x_1+x_2\,,
\label{om-eff}\\
w_\eff &= \frac{P_\eff}{\rho_\eff}\nonumber\\
&= \frac{1}
{\left(6 x_4+3\right)\Omega_\eff}\left( 2 x_4 \left(2 x_4 \left(4 \epsilon  h_N+2 h_N+4 \epsilon _N
+4 \epsilon ^2+8 \epsilon +3\right)
\right. \right.
\nonumber\\
&\left. \left.
+4 \epsilon  h_N+2 h_N+4 \epsilon _N+16 \epsilon ^2+8 \epsilon +3\right) -3 \left(2 x_4+1\right) x_1+x_2 \left(2 x_4+1\right) \right)\,,
\label{w-eff}
\end{align}
where $\Omega_\eff$ is the density parameter, $w_\eff$ is the effective equation of state parameter,
and $\epsilon_N \equiv d \epsilon / dN$. 
The derivative of $\epsilon$ is obtained by differentiating Eq.~(\ref{epsilon}) with respect to $N$,
and use Eqs.~(\ref{x1n})--(\ref{omn}) to write the result in terms of the dimensionless variables.

\section{Reduction of equations}
\label{app:reduction}

In this appendix, we outline the procedure used to eliminate higher-order derivatives from the background evolution equations.
Combining Eqs.~(\ref{e00}) and (\ref{eii}), $\dotH$ and $\dddot{\phi}$ can be eliminated to give
\begin{align}
\frac{12}{a G_4} \left(G_4 X \left(-6 G_4 H^2 G_{4 X}-6 H \dot{\phi } G_{4 X}^2 \ddot{\phi }+6 G_4 H \dot{\phi } G_{3 X}+3 G_{4 X}^2 \ddot{\phi }^2-2 G_4 G_{2 X}+3 G_2 G_{4 X}\right) \right. \nonumber\\
\left. +G_4^2 \left(6 H \left(G_4 H+\dot{\phi } G_{4 X} \ddot{\phi }\right)+G_2\right)-G_4 \rho _\M \left(G_4-3 X G_{4 X} w_\M\right) \right. \nonumber\\
\left.-3 X^2 G_{4 X} \ddot{\phi } \left(G_{4 X}^2 \ddot{\phi }-2 G_4 G_{3 X}\right)\right)
 = 0\,.
 \label{e1}
\end{align}
Differentiating Eq.~(\ref{eii}) with respect to time, we obtain an equation that depends on $\ddddot{\phi}$ and $\ddot{H}$.
This equation can be used to eliminate $\ddddot{\phi}$ and $\ddot{H}$ from Eq.~(\ref{kg}).
The resulting equation depends on $\dddot{\phi}$ and $\dotH$, which can be eliminated using Eq.~(\ref{eii}) to give
\begin{align}
\frac{36 X G_{4 X}^2 }{G_4^3} \left(6 G_4^2 H^2 \dot{\phi } \left(X \left(\left(G_{4 X}^2-2 G_4 G_{4 X X}\right) \ddot{\phi }+3 G_4 G_{3 X}\right)-G_4 G_{4 X} \ddot{\phi }\right) \right. \nonumber\\
\left.+6 G_4 H X \left(G_4 G_{4 X} \rho _\M \left(w_{\M N}-3 w_\M^2+w_\M\right) \right. \right. \nonumber\\
\left. \left.
+2 X \ddot{\phi } \left(G_{4 X}^3 \ddot{\phi }-2 G_4 G_{4 X X} G_{4 X} \ddot{\phi }+2 G_4 G_{3 X} G_{4 X}+2 G_4^2 G_{3 X X}\right) \right. \right. 
\nonumber\\
\left. \left.
-2 G_4 \left(G_{4 X}^2 \ddot{\phi }^2-2 G_4 G_{3 X} \ddot{\phi }+G_4 G_{2 X}-2 G_2 G_{4 X}\right)\right) \right.
\nonumber\\
\left. 
+\dot{\phi } \left(-G_4 X \left(3 \rho _\M w_\M \left(\left(G_{4 X}^2-2 G_4 G_{4 X X}\right) \ddot{\phi }+G_4 G_{3 X}\right) \right. \right. \right.
\nonumber\\
\left. \left. \left.
+3 G_2 \left(\left(G_{4 X}^2-2 G_4 G_{4 X X}\right) \ddot{\phi }+G_4 G_{3 X}\right)  \right. \right. \right.
\nonumber\\
\left. \left. \left.
+\ddot{\phi } \left(3 G_{4 X} \ddot{\phi } \left(G_{4 X}^2 \ddot{\phi }-4 G_4 G_{3 X}\right)+4 G_4^2 G_{2 X X}-6 G_4 G_{2 X} G_{4 X}\right)\right) \right. \right.
\nonumber\\
\left. \left.
+G_4^2 \ddot{\phi } \left(3 G_{4 X} \rho _\M w_\M-2 G_4 G_{2 X}+3 G_2 G_{4 X}\right) \right. \right.
\nonumber\\
\left. \left.
+3 X^2 \ddot{\phi } \left(\left(G_{4 X}^4-2 G_4 G_{4 X}^2 G_{4 X X}\right) \ddot{\phi }^2 \right. \right. \right.
\nonumber\\
\left. \left. \left.
+G_4 G_{4 X} \left(G_{3 X} G_{4 X}+4 G_4 G_{3 X X}\right) \ddot{\phi }-2 G_4^2 G_{3 X}^2\right)\right)\right) = 0\,.
\label{e2}
\end{align}
The quadratic equation of $\ddot{\phi}$ is obtained by combining Eqs.~(\ref{e1}) and (\ref{e2}) as
\be
A \ddot\phi^2 + B \ddot\phi + C = 0\,,
\label{eq-ddp}
\ee
where expressions of $A$, $B$, and $C$ are 
\begin{align}
A &= -\frac{432 X^2 \dot{\phi } G_{4 X}^3 \left(X^2 \left(G_{3 X} G_{4 X X}-G_{4 X} G_{3 X X}\right)+X \left(G_4 G_{3 X X}-G_{3 X} G_{4 X}\right)+G_4 G_{3 X}\right)}{G_4 \left(G_4-X G_{4 X}\right)}\,,
\label{defA}\\
B &= \frac{36 X G_{4 X}^2}{G_4 \left(G_4-X G_{4 X}\right)} \left(\dot{\phi } \left(2 \left(6 X^3 G_{3 X}^2 G_{4 X}+X^2 \left(3 G_4 G_{3 X}^2-2 G_4 G_{4 X} G_{2 X X} \right. \right. \right. \right.
\nonumber\\
&\left. \left. \left. \left. 
+2 G_{2 X} \left(G_{4 X}^2+G_4 G_{4 X X}\right)\right)+X \left(2 G_4^2 G_{2 X X}-3 G_4 G_{2 X} G_{4 X} \right. \right. \right. \right.
\nonumber\\
&\left. \left. \left. \left.
+2 G_2 \left(G_{4 X}^2-2 G_4 G_{4 X X}\right)\right)+G_4 \left(G_4 G_{2 X}-2 G_2 G_{4 X}\right)\right) \right. \right.
\nonumber\\
&\left. \left.
-\left(G_4 G_{4 X}-X \left(G_{4 X}^2-2 G_4 G_{4 X X}\right)\right) \rho _\M \left(3 w_\M-1\right)\right) \right.
\nonumber\\
&\left.
-24 G_4 H X \left(X^2 \left(G_{3 X} G_{4 X X}-G_{4 X} G_{3 X X}\right)+X \left(G_4 G_{3 X X}-G_{3 X} G_{4 X}\right)+G_4 G_{3 X}\right)\right)\,,
\label{defB}\\
C &=  \frac{108 X^2 G_{4 X}^2}{G_4 \left(G_4-X G_{4 X}\right)} \left(2 H \left(G_{4 X} \left(G_4-X G_{4 X}\right) \rho _\M \left(-w_{\M N}+3 w_\M^2-w_\M\right)+6 G_4 X^2 G_{3 X}^2 \right. \right.
\nonumber\\
&\left. \left.
+2 G_4 \left(G_4 G_{2 X}-2 G_2 G_{4 X}\right)+2 X G_{4 X} \left(2 G_2 G_{4 X}-G_4 G_{2 X}\right)\right) \right.
\nonumber\\
&\left.
+\dot{\phi } G_{3 X} \left(-G_4 \rho _\M+\left(2 X G_{4 X}+G_4\right) \rho _\M w_\M+2 X \left(G_2 G_{4 X}-G_4 G_{2 X}\right)+2 G_2 G_4\right)\right)\,.
\label{defC}
\end{align}
The expression for $\dotH / H^2$ can be computed by differentiating Eq.~(\ref{e2}) with respect to time,
and the term $\dddot\phi$ can be eliminated using Eq.~(\ref{eii}), yielding an equation of the form
\be
\frac{\dotH}{H^2} \equiv h_N(\ddot\phi, \dot\phi, H, \rho_\M, w_\M)\,.
\label{dh2h2}
\ee
The solutions for Eq.~\eqref{eq-ddp} are
\be
\epsilon = a_1 \pm \sqrt{a_1^2 + a_2}\,,
\ee 
where 
\begin{align}
a_1 &=\frac{1}{96 x_3 x_4 \left(2 x_4+1\right)} \left( x_1 \left(12 x_4^2-24 x_4+1\right)-6 \left(-2 x_4 \left(2 x_4+3\right) \Omega _m \left(3 w_M-1\right)+\left(10 x_4+1\right) x_3^2 \right.\right.\nonumber \\ 
& \left.\left.+2 \left(2 x_4+1\right)^2 x_3+2 x_2 \left(1-2 x_4\right)^2\right) \right), 
\label{defA1}\\
a_2 &=
\frac{3}{16 x_3 x_4 \left(2 x_4+1\right)} \left( \Omega _m \left(4 x_4 \left(2 x_4-1\right) \left(-w_{M N}+3 w_M^2-w_M\right)+x_3 \left(10 x_4 w_M+w_M-2 x_4-1\right)\right)
\right.\nonumber \\ 
& \left.
+\left(12 x_4^2-4 \left(x_3+2\right) x_4+1\right) x_1+2 x_2 \left(x_3-4 x_4+2\right) \left(2 x_4-1\right)-6 x_3^2 \left(2 x_4+1\right) \right)\,.
\label{defA2}
\end{align}

\section{Approximations of $\epsilon$, $h_N$, $\epsilon_N$, $\Omega_\eff$, and $w_\eff$ in terms of dimensionless variables}
\label{app:approx:dimless}
 
During the matter dominated epoch, the approximations of $\epsilon$, $h_N$, $\epsilon_N$, $\Omega_\eff$, and $w_\eff$ in terms of the dimensionless variables are given by
\begin{align}
\label{epsilon-mat}
\epsilon & \simeq -\frac{3 \left(x_1^2-4 x_2\right)}{x_1^2-12 (3 \Omega_m x_4+x_2)} \simeq 0\,, \\
h_N & \simeq -\frac{3}{2} + \frac{1}{4 \left(x_1^2-12 (x_2+3 x_4)\right) \left(x_1^2-12 (3 \Omega_m x_4+x_2)\right)^2}\left( 24 x_1^4 \left(18 \left(3 \Omega_m^2+106 \Omega_m-92\right) x_4^2 \right. \right.  \nonumber \\ 
& \left. \left. +(78 \Omega_m+395) x_2 x_4+21 x_2^2\right)-288 x_1^2 \left(3 \left(21 \Omega_m^2+506 \Omega_m-448\right) x_2 x_4^2+4 (12 \Omega_m+49) x_2^2 x_4 \right. \right.  \nonumber \\ 
& \left. \left. +54 \Omega_m (9 \Omega_m-20) x_4^3+9 x_2^3\right)+3456 x_2 \left(3 \left(3 \Omega_m^2+94 \Omega_m-88\right) x_2 x_4^2+(6 \Omega_m+31) x_2^2 x_4
\right. \right.  \nonumber \\ 
& \left. \left.
+45 \Omega_m (3 \Omega_m-8) x_4^3+x_2^3\right)-2 x_1^6 (6 (6 \Omega_m+49) x_4+19 x_2)+x_1^8 \right) \simeq -\frac{3}{2}\,,
\\
\epsilon_N & \simeq \frac{36 \left(x_1^2-4 x_2\right) \left(x_1^2 (27 \Omega_m x_4+4 x_2)-108 \Omega_m x_4 (3 \Omega_m x_4+x_2)\right)}{\left(x_1^2-12 (3 \Omega_m x_4+x_2)\right)^3} \simeq 0 \,,\label{depsilon-mat} \\
\Omega_\eff & \simeq \frac{-6 x_1^2 (6 \Omega_m x_4+3 x_2+26 x_4)+72 \left((3 \Omega_m+10) x_2 x_4+6 \Omega_m x_4^2+x_2^2\right)+x_1^4}{72 (3 \Omega_m x_4+x_2)-6 x_1^2}\,,
\label{omphi}\\ 
w_\eff & \simeq  \left( -288 x_1^2 \left(54 \Omega_m^2 (3 \Omega_m-14) x_4^3+4 (18 \Omega_m+43) x_2^2 x_4+3 \Omega_m (63 \Omega_m+16) x_2 x_4^2+9 x_2^3\right) \right. 
\nonumber \\  
&\left.  +3456 x_2 \left(27 \Omega_m^2 x_2 x_4^2+9 \Omega_m^2 (3 \Omega_m-28) x_4^3+(9 \Omega_m+28) x_2^2 x_4+x_2^3\right)+24 x_1^4 \left((117 \Omega_m+356) x_2 x_4 \right. \right.
\nonumber \\  
&\left. \left. + 18 \Omega_m (9 \Omega_m+8) x_4^2+21 x_2^2\right)-2 x_1^6 (6 (9 \Omega_m+46) x_4+19 x_2)+x_1^8 \right) /
\nonumber \\  
&\left( (x_1^2-12 (3 \Omega_m x_4+x_2))^2 \left(-6 x_1^2 (6 \Omega_m x_4+3 x_2+26 x_4)+72 \left((3 \Omega_m+10) x_2 x_4+6 \Omega_m x_4^2+x_2^2\right) 
\right. \right.
\nonumber \\  
&\left. \left. +x_1^4\right) \right).
\label{wphi}
\end{align}

\section{Expressions for the perturbed Lagrangian}
\label{app:pert}

The expressions for $\call_2$, $\call_3$, and $\call_4$ are given by
\ba
\label{lp2}
\call_2 &&= -\Phi \delta\rho_m - \frac{1}{2} a P_X \pi {}_{,i} \pi {}^{,i}
+ \frac{a}{\dph^4} \Bigl(2 \dph^4 G_{4}\bigl(-3 \beta^2 \Phi + 2 \beta (\Phi - 2 \Psi) + 2 \Psi \bigr) \Phi {}^{,i}{}_{,i} - 24 \beta^2 \dph^3 G_{4}H \Phi \pi {}^{,i}{}_{,i}
\nonumber \\ &&- 2 \dph^4 G_{4}\Psi \Psi {}^{,i}{}_{,i} + 4 \beta (-1 + 3 \beta) \dph^3 G_{4}\Phi \dot\pi_{,i}^{,i} + 8 \beta \dph^3 G_{4}\Psi \dot\pi_{,i}^{,i} - 6 \beta^2 \ddph^2 G_{4}\pi {}_{,i} \pi {}^{,i} + 6 \beta^3 \ddph^2 G_{4}\pi {}_{,i} \pi {}^{,i} 
\nonumber \\ &&
- 6 \beta^2 \dddph \dph G_{4}\pi {}_{,i} \pi {}^{,i} 4
- 3 \beta \ddph^2 \dph^4  G_{4XX} \pi {}_{,i} \pi {}^{,i} + 6 \beta^2 \ddph \dph G_{4}H \pi {}_{,i} \pi {}^{,i} - 24 \beta^3 \ddph \dph G_{4}H \pi {}_{,i} \pi {}^{,i} 
\nonumber \\ &&
+ 12 \beta \ddph \dph^5  G_{4XX} H \pi {}_{,i} \pi {}^{,i} + 6 \beta \dph^2 G_{4}H^2 \pi {}_{,i} \pi {}^{,i} + 36 \beta^2 \dph^2 G_{4}H^2 \pi {}_{,i} \pi {}^{,i} + 12 \beta^2 \dotH \dph^2 G_{4}H^2 \pi {}_{,i} \pi {}^{,i} 
\nonumber \\ &&
- 12 \beta \ddph \dph^2 G_{4}^{1/2} \bar\rho_m \pi_{,i} \pi^{,i} + 24 \beta \dph^3 G_{4}^{1/2} H \bar\rho_m \pi_{,i} \pi^{,i} + 6 \beta^2 \dph^2 G_{4}\dot\pi_{,i} \dot\pi^{,i}\Bigr)\,,
\\
\call_3 &&= \frac{2}{a \dph^3 G_{4}}  \pi^{,i} \Bigl(\beta G_{4}G_{4} \pi {}_{,i} \bigl((\dph - 3 \beta \dph) \Phi {}^{,j}{}_{,j} - 2 \dph \Psi {}^{,j}{}_{,j} + 3 \beta \dot\pi^{,j}_{,j}\bigr) + 3 \dph^4  G_{4X}^2 H \pi_{,i ,j} \pi^{,j}\Bigr)\,,
\label{lp3} \\
\call_4 &&= \frac{6\beta^2 G_{4}}{a^3 \dph^4}\pi^{,i} \pi_{,k ,j} \pi^{,j} \pi_{,i}^{,k}\,.
\label{lp4}
\ea
In the above Lagrangians, we define $\beta \equiv G_{4X} \dot{\phi} / (2 G_4)$, and ${}_{,i}$ denotes a derivative with respect to the spatial comoving coordinates.

\section{Equations of motion for the perturbed variables}
\label{app:perteom:all} 

The equations of motion for $\Phi$, $\Psi$, and $\pi$ are given by
\ba
\label{eom:all}
- a^2 \delta \dph 
&=& 4 (1 - 3 \beta) \beta G_{4}\dpi^{, i}_{, i} + 4 G_{4}\bigl(3 \beta^2 (\dph \Phi^{, i}_{, i} + 2 H \pi^{, i}_{, i}) -  \dph \Psi^{, i}_{, i} + 2 \beta \dph (- \Phi^{, i}_{, i} + \Psi^{, i}_{, i})\bigr)
\nonumber\\
&& - \frac{2 (1 - 3 G_{4}) G_{4}\beta}{a^2 \dot\phi}\Pi^{,i}_{,i}\,,
\label{eomphi}\\
0 &=& \frac{2 \beta \dpi^{, i}_{, i}}{\dph} + \Phi^{, i}_{, i} - 2 \beta \Phi^{, i}_{, i} - \Psi^{, i}_{, i}
\nonumber\\
&& - \frac{\beta}{a^2 \dot\phi^2}\Pi^{,i}_{,i}\,,
\label{eompsi}
\ea 
\ba
0 &=& \frac{a}{2 \dph^4} \Bigl(8 \beta (-1 + 3 \beta) \dph^3 \dPh G_{4}+ 48 \beta^2 \ddph \dph \dot\pi^{, i}_{,i} G_{4}+ 16 \beta \dph^3 \dPs G_{4} - 24 \beta^2 \ddph \dph^3 \dot\pi^{, i}_{,i}  G_{4X} - 96 \beta^2 \dph^2 \dot\pi^{, i}_{,i} G_{4}H 
\nonumber \\ &&
+ 8 \beta \ddph \dph^2 G_{4}\Phi 
- 24 \beta^2 \ddph \dph^2 G_{4}\Phi - 8 \beta \ddph \dph^4  G_{4X} \Phi + 24 \beta^2 \ddph \dph^4  G_{4X} \Phi - 32 \beta \dph^3 G_{4}H \Phi + 144 \beta^2 \dph^3 G_{4}H \Phi 
\nonumber \\ 
&&
- 24 \beta^2 \ddph^2 G_{4}\pi^{, i}_{,i} + 24 \beta^3 \ddph^2 G_{4}\pi^{, i}_{,i} - 24 \beta^2 \dddph \dph G_{4}\pi^{, i}_{,i} - 12 \beta \ddph^2 \dph^4  G_{4XX} \pi^{, i}_{,i} + 24 \beta^2 \ddph \dph G_{4}H \pi^{, i}_{,i} 
\nonumber \\ 
&&
- 96 \beta^3 \ddph \dph G_{4}H \pi^{, i}_{,i} 
+ 48 \beta \ddph \dph^5  G_{4XX} H \pi^{, i}_{,i} + 24 \beta \dph^2 G_{4}H^2 \pi
+ 144 \beta^2 \dph^2 G_{4}H^2 \pi^{, i}_{,i} + 48 \beta^2 \dph^2 G_{4}\dotH \pi^{, i}_{,i} \nonumber \\ 
&& - 16 \beta \ddph \dph^2 G_{4}\Psi
+ 16 \beta \ddph \dph^4  G_{4X} \Psi + 64 \beta \dph^3 G_{4}H \Psi -  \dph^4 \pi^{, i}_{,i} P_X + 8 \dph^2 G_{4}\bigl(-6 \beta \dot\pi^{, i}_{,i} + (-1 + 6 \beta) \dph \Phi \nonumber \\ 
&&
+ 2 \dph \Psi \bigr) \dot{\beta}
- 24 \beta^2 \dph^2 G_{4}\ddot \pi^{, i}_{,i} \Bigr)
- \frac{2}{a \beta \dot\phi^4}
\Bigl(9 G_{4}^2 \beta^2 \ddot\phi \Pi^{, i}_{,i} - 3 G_{4}\beta \dot\phi \bigl(G_{4} \beta \dot\Pi^{, i}_{,i} + 2 G_{4}\beta \dot\pi^{,j}_{,j} \pi^{, i}_{,i} + (- G_{4}\beta H 
\nonumber \\ 
&&
+ 2 \beta \dot{G_{4}}
+ G_{4}\dot\beta) \Pi^{, i}_{,i}\bigr) 
+ 2 G_{4}\beta^2 \dot\phi^2 \pi^{, i}_{,i} \bigl((-1 + 3 G_{4}) \Phi^{,j}_{,j} + 2 \Psi^{,j}_{,j}\bigr) 
+ 3  G_{4X}^2 H \dot\phi^5 (\pi^{, i}_{,i} \pi^{,j}_{,j} -  \pi^{, i}_{,i, j} \pi^{,j})\Bigr)
\nonumber\\ 
&& - \frac{12 G_{4}^2 \beta}{a^3 \dot\phi^4} 
\pi^{, i}_{,i} (\pi^{,j} \pi^{,k}_{,j ,k} + \pi_{,j ,k} \pi^{,j ,k})\,,
\label{eomfield}
\ea
where $\Pi \equiv \pi_{, i} \pi^{, i}$.

\section{Linear perturbations}
\label{app:linpert}

We derive the linear perturbation equations under the quasi-static approximation.
Eqs.~(\ref{eomphi})--(\ref{eomfield}) can be solved to obtain $\Phi$ and $\Psi$ as
\ba
\label{eom:all:redu}
\Phi_{, i}^{, i} &=& \frac{a^2 \delta \dph + 24 \beta^2 G_{4}H \pi^{, i}_{,i} - 4 \beta G_{4}\dot\pi^{, i}_{,i} + 4 \beta^2 G_{4}\dot\pi^{, i}_{,i} }{4 (-1 + \beta)^2 \dph G_{4}}\,,
\\
\Psi_{, i}^{, i} &=& \frac{a^2 \delta \dph - 2 a^2 \beta \delta \dph + 24 \beta^2 G_{4}H \pi^{, i}_{,i} - 48 \beta^3 G_{4}H \pi^{, i}_{,i} + 4 \beta G_{4}\dot \pi^{, i}_{,i} - 4 \beta^2 G_{4}\dot \pi^{, i}_{,i}}{4 (-1 + \beta)^2 \dph G_{4}}\,. 
\label{psisol}
\ea
Differentiating Eqs.~(\ref{eom:all:redu}) and (\ref{psisol}) with respect to time,
and substituting the resulting expressions for $\dot{\Phi}$ and $\dot{\Psi}$, together with Eqs.~(\ref{eom:all:redu}) and (\ref{psisol}), into Eq.~(\ref{eomfield}), we obtain
\be
\pi_{, i}^{, i} = - \frac{a^2 \dph^2}{G_{4} {\cal G} } 
\Bigl(\delta \bigl(2 \beta^2 \ddph G_{4}^2 + \ddph \dph^2 (2 G_{4}G_{4X} -  \dph^2  G_{4X}^2 + \dph^2 G_{4}G_{4XX}) - 2 \beta G_{4}^2 (\ddph - 6 \dph H)\bigr) - 2 (-1 + \beta) \beta \dph G_{4}^2 \dot \delta \Bigr)\,, 
\label{pisol}
\ee
where
\ba
{\cal G} &&\equiv 24 \beta^5 \ddph G_{4}^2 (\ddph - 4 \dph H) - 24 \beta^4 G_{4}\bigl(3 \ddph^2 G_{4}+ \ddph \dph (-13 G_{4}+ 2 \dph^2  G_{4X}) H + \dph G_{4}(\dddph - 10 \dph H^2)\bigr)
\nonumber \\ 
&&
+ 12 \beta^3 \bigl(\ddph^2 G_{4}(6 G_{4}
-  \dph^4  G_{4XX}) + 4 \ddph \dph (-5 G_{4}^2 -  \dph^2 G_{4}G_{4X} + \dph^4  G_{4X}^2) H - 2 \dph G_{4}^2 (-2 \dddph + 3 \dph H^2 + 2 \dph \dotH)\bigr) 
\nonumber \\ 
&&
-  \dph^4 G_{4}P_X 
+ 2 \beta \dph^2 G_{4}(-6 \ddph^2 \dph^2  G_{4XX} + 24 \ddph \dph^3  G_{4XX} H + 12 G_{4}H^2 + \dph^2 P_X) + \beta^2 \Bigl(-24 \ddph^2 G_{4}(G_{4} 
\nonumber \\ 
&&
-  \dph^4  G_{4XX}) - 24 \ddph \dph (- G_{4}^2 - 6 \dph^2 G_{4}G_{4X} + 3 \dph^4  G_{4X}^2 + \dph^4 G_{4}G_{4XX}) H + \dph G_{4}\bigl(-24 \dddph G_{4}+ 48 \dph G_{4}(2 H^2 
\nonumber \\ 
&&
+ \dotH) -  \dph^3 P_X\bigr)\Bigr)\,.
\ea
Differentiating Eq.~(\ref{pisol}) with respect to time and substituting both Eq.~(\ref{pisol}) and its time derivative into Eq.~(\ref{psisol}), we obtain
\be
\Psi_{, i}^{, i} = C_1 \delta_m + C_2 \dot \delta_m + C_3 \ddot\delta_m\,, 
\label{psisolF}
\ee
where the coefficients $C_1$, $C_2$, and $C_3$ are functions of $\dot\phi$, $\ddot\phi$, $\dddot\phi$, $H$, and $\dotH$, with lengthy explicit forms omitted.
\section{Nonlinear collapse equations}
\label{app:collapse}

In this section, we derive the modified Poisson equation used in the spherical collapse analysis.
To compute the modified Poisson equation,
we assume the spherical symmetry and use the spherical coordinates with Eq.~\eqref{ds2pert}.
A spherically symmetric distribution of matter yields $\rho=\rho(t,r)$, where $r$ is the comoving radial coordinate. 
We follow \cite{Hirano:1903} to write The perturbation variables in terms of dimensionless quantities as
\begin{align}
&x \equiv \frac{\pi_{,r}}{\Lambda^3 r},\quad
y \equiv \frac{G_4^{1/2}\Phi_{,r}}{\Lambda^3 r},\quad 
z \equiv \frac{G_4^{1/2}\Psi_{,r}}{\Lambda^3 r},\quad 
\beta \equiv \frac{\dot\phi^2 G_{4X} X}{2 G_{4}},
\\
&A \equiv \frac{1}{8\pi \dot\phi^2}\frac{M(t,r)}{r^3}\,,
\label{defA}
\end{align}
where $\Lambda^3 \equiv \dot{\phi}^2_0 / G_4^{1/2}$,
$\dot\phi_0$ is the present value of the time-derivative of the field, 
and the mass contained within radius $r$ is given by
\begin{align}
M(t,r) \equiv 4\pi \int^r_0 \delta\rho_m(t, \tilde{r}) \tilde{r}^2 d\tilde{r}\,.
\label{def:mass}
\end{align}
The Vainshtein radius can be defined as
\be
r_* \equiv \(\frac{M(t,r)}{8\pi \dot\phi^2}\)^{1/3}\,,
\ee
so that
\be
A = \(\frac{r_*}{r}\)^3\,.
\ee
In terms of the dimensionless variables,
Eqs.~\eqref{eomphi}--\eqref{eomfield} become
\begin{align}
0 &= - a^2 A + \dph (-2 + 3 \beta) \Lambda^3 y + (2 - 4 \beta) z 
\nonumber \\
&\quad - \left( \frac{6 \dot\lm}{\dph \Lambda} \beta^{3/2} (1 - 3 \beta) x
+ \frac{2}{\dph} \beta^{3/2} \dot{x} 
+ \frac{12}{\dph} H \beta^{5/2} x 
- \frac{2}{\dph} 3 \beta^{5/2} \dot{x} \right)
\nonumber \\
&\quad - \frac{2}{a^2 \dph^2} \beta^{3/2} (-1 + 3 \beta) \Lambda^3 x (x + r x_{,r}) \,,
\label{phieq}\\ 
0 &= (1 - 2 \beta) y - z  
\nonumber \\
&\quad + \frac{2}{a^2 \dph^2} \beta^{3/2} \left( a^2 \dph \left(3 \frac{\dot\lm}{\lm} x + \dot{x} \right) - x (x + r x_{,r}) \Lambda^3 \right)\,,
\label{psieq}
\end{align}
and
\begin{align}
{\cal F}(x,\dot x,x_{,r},\ddot x,\dot x_{,r},x_{,rr},y,\dot y, y_{,r}, z, \dot z, z_{,r})=0\,,
\label{pieq}
\end{align}
where ${}_{,r}$ denotes differentiation with respect to $r$, and the explicit form of ${\cal F}$ is complicated.
Here, the linear perturbation terms that depend on both time and spatial derivatives (such as $\dot{x}$) cannot be neglected; otherwise, the nonlinear equations would not reduce to the linear equations obtained in the previous section.
Eqs.~\eqref{phieq} and \eqref{psieq} can be solved for $y$ and $z$. Their expressions are
\begin{align}
y &= \frac{a^4 \dph^2 A \Lambda^3 + 2 \beta^{3/2} \Bigl(- x (x + r x_{,r}) (-1 + \beta) \Lambda^6 + a^2 \dph \bigl(3 \lm^2 \dot\lm x (-1 + \beta) + \dot{x} (-1 + \beta) \Lambda^3 + 6 H x \beta \Lambda^3\bigr)\Bigr)}{2 a^2 \dph^2 (-1 + \beta)^2 \Lambda^3},
\label{eq:y=}\\
z &= \frac{1}{2 a^2 \dph^2 (-1 + \beta)^2 \Lambda^3} \Bigl( a^4 \dph^2 A (1 - 2 \beta) \Lambda^3 + 2 \beta^{1/2} x (x + r x_{,r}) (-1 + \beta) \beta \Lambda^6 
\nonumber \\
&\quad - 2 a^2 \dph \beta^{1/2} \beta \bigl(3 \lm^2 \dot\lm x (-1 + \beta)
+ \dot{x} (-1 + \beta) \Lambda^3 + 6 H x \beta (-1 + 2 \beta) \Lambda^3\bigr) \Bigr),
\label{eq:z=}
\end{align}
Differentiating Eqs.~\eqref{eq:y=} and \eqref{eq:z=} with respect to $t$ and $r$ to obtain $\dot{y}$, $\dot{z}$, $y_{,r}$, and $z_{,r}$, and substituting these into Eq.~\eqref{pieq}, we obtain
\begin{align}
\al x^2 + \left(\bt_1 + \frac{\bt_2}{3 r^2} \frac{d}{dr}(r^3 A)\right) x + \mu A + \nu \dot{A} = 0\,,
\label{eq:x20}
\end{align}
Applying the spherical top-hat assumption, Eq.~\eqref{defA} yields
\be
A = \frac{\delta\rho_m}{6 \dot\phi^2}\,,
\label{a-tophat}
\ee
and consequently $A_{,r} = x_{,r} = 0$.
Hence, under the top-hat assumption, Eq.~\eqref{eq:x2} becomes
\begin{align}
\al x^2 + \left(\bt_1 + \bt_2 A\right) x + \mu A + \nu \dot{A} = 0\,,
\label{eq:x2}
\end{align}
The coefficients in Eq.~\eqref{eq:x2} are
\ba
\al &=& -3 \dph^4 G_{4X} (2 \beta -  \dph^2 G_{4X}) \bigl(\ddph (2 \beta -  \dph^2 G_{4X}) (5 \beta G_{4X} - 2 \dph^2 G_{4X}^2 + 2 \dph^2 \beta G_{4XX}) + \dph \beta G_{4X} (26 \beta
\nonumber \\&&
- 5 \dph^2 G_{4X}) H\bigr)\,,
\\
\bt_1 &=& - a^2 \dph^2 \beta^{1/2} \biggl(3 \ddph^2 G_{4X} (-2 \beta + \dph^2 G_{4X})^2 \bigl(- \dph^2 G_{4X}^2 + 2 \beta (G_{4X} + \dph^2 G_{4XX})\bigr) + \cdots \biggr)\,,
\\
\bt_2 &=& 12 a^2 \dph^4 \beta^{5/2} G_{4X} (-2 \beta + \dph^2 G_{4X})\,,
\\
\mu &=& 12 a^4 \dph^2 \beta^3 \bigl(\ddph (6 \beta G_{4X} - 3 \dph^2 G_{4X}^2 + 2 \dph^2 \beta G_{4XX}) + 12 \dph \beta G_{4X} H\bigr)\,,
\\
\nu &=& -12 a^4 \dph^3 \beta^3 G_{4X} (-2 \beta + \dph^2 G_{4X})\,.
\ea
In spherical symmetry, we have $\Psi^{, i}_{, i} = \dot\phi^2 (r^3 z)_{,r} / G_4$.
Hence, the Poisson equation $\Psi^{, i}_{, i}$ can be obtained by applying the operator $\bD$ to Eq.~\eqref{eq:z=}, where
\begin{align}
\bD \chi \equiv \frac{1}{r^2}\frac{\partial}{\partial r}(r^3 \chi)\,.
\label{opea}
\end{align}
Here, $\chi$ denotes any perturbation variable, such as
\be
\bD x = \frac{1}{\Lambda^3} \nabla^2\pi,\quad
\bD z = \frac{G_4}{\dot\phi^2} \nabla^2\Psi,\quad
\bD A = \frac{1}{2 \dot\phi^2}\delta\rho_m\,.
\ee
Applying $\bD$ to Eq.~\eqref{eq:z=} yields
\ba
\nabla^2\Psi &= \frac{a^2}{4 G_4 (-1 + \beta)} \delta\rho_m 
+ \frac{\beta^{3/2} \dot\phi^2}{a^2 G_4^{3/2} (-1 + \beta)} \bD x^2 
\nonumber \\
&\quad - \frac{\beta^{3/2}}{\dph (-1 + \beta)^2 G_4^{1/2}}
\left[ \left( \dot{\Lambda}^3 + 6 H \beta (-1 + 2 \beta) \Lambda^3 \right) \bD x
+ (-1 + \beta) \Lambda^3 \frac{d \bD x}{dt} \right]\,.
\label{pos0}
\ea
Using $x_{,r} = 0$, we obtain
\be
(\bD x)^2 = 9 x^2,\quad
\bD x^2 = 3 x^2\,,
\ee
so that
\be
\bD x^2 = \frac{1}{3} (\bD x)^2\,.
\label{dx2TOdx2}
\ee
Defining $\tilde{x} \equiv \bD x$, Eq.~\eqref{pos0} can be written as
\be
\nabla^2\Psi = g(\delta\rho_m, \dot\delta\rho_m, \tilde{x}, \tilde{x}^2, \dot{\tilde{x}})\,.
\ee
To express $\tilde{x}$ in terms of $\delta\rho_m$, we apply $\bD$ to Eq.~\eqref{eq:x2} and obtain
\begin{align}
\frac{\al}{3} \tilde{x}^2 + \left(\bt_1 + \frac{\bt_2}{2 \dot\phi^2}\delta\rho_m\right) \tilde{x} + \frac{\mu}{2 \dot\phi^2}\delta\rho_m + \frac{d}{dt}\left[\frac{\nu}{2 \dot\phi^2}\delta\rho_m\right] = 0\,,
\label{eq:x2-v1}
\end{align}
which yields
\be
\tilde{x}_\pm = \frac{3}{2 \al}\left(- B \pm \sqrt{B^2 - \frac{4}{3} \al C}\right)\,,
\label{xxsol}
\ee
where
\be
B \equiv \left(b_1 + \frac{b_2}{2 \dot\phi^2}\delta\rho_m\right),\quad
C \equiv \frac{\mu}{2 \dot\phi^2}\delta\rho_m + \frac{\nu}{dt}\left[\frac{d}{2 \dot\phi^2}\delta\rho_m\right]\,.
\ee
We select the plus-branch solution in Eq.~\eqref{xxsol}, since the minus branch leads to solutions dominated by background quantities in the linear regime $\delta\rho_m / \bar\rho_m \ll 1$.

\begin{figure}[htbp]
\centering
\includegraphics[width=0.6\textwidth, angle=0]{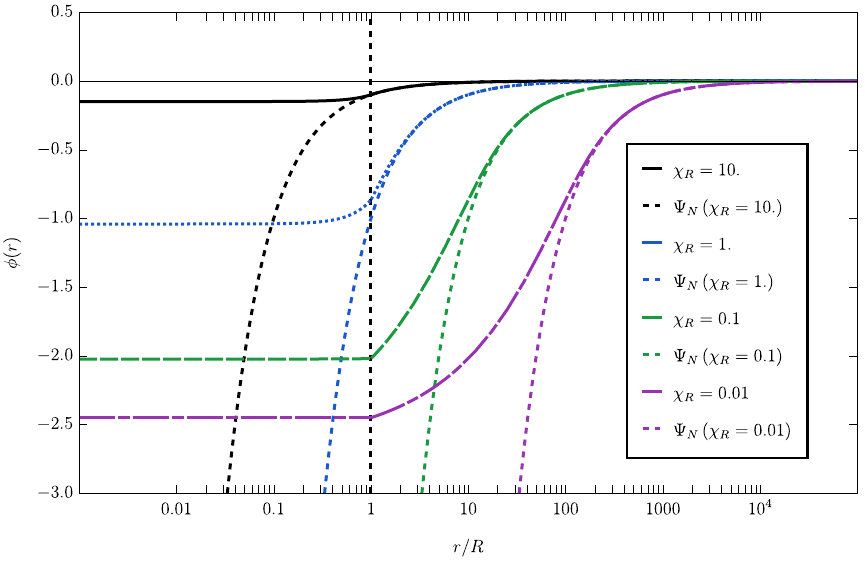}
\caption{The field profile as a function of the normalized radial distance $r/R$ for various values of the source parameter $\chi_R$, compared with the corresponding Newtonian potentials $\Psi_N$ (dashed lines). }
\label{fig:fp}
\end{figure}

Substituting $\tilde{x}_+$ into the Poisson equation, we obtain
\ba
\nabla^2\Psi &= & \frac{3 a^2 H^2 \Om_m}{4 G_4 (-1 + \beta)} \delta_m 
+ \frac{\beta^{3/2} \dot\phi^2}{a^2 G_4^{3/2} (-1 + \beta)} \tilde{x}_+^2
\nonumber \\
&&\quad - \frac{\beta^{3/2}}{\dph (-1 + \beta)^2 G_4^{1/2}}
\left[ \left( \dot{\Lambda}^3 + 6 H \beta (-1 + 2 \beta) \Lambda^3 \right) \tilde{x}_+  
+ (-1 + \beta) \Lambda^3 \dot{\tilde{x}}_{+}\right]\,.
\label{pos1}
\ea
Substituting the above expression into Eq.~\eqref{eq:collapseeqn} and keeping $(1+\de)\nabla^2\Psi$ up to second order in perturbations, we finally obtain
\be
\de_{m\,NN} + \left(2 + \frac{\dotH}{H^2}\right)\de_{m\, N} 
- \frac{4}{3}\frac{\de_{m\, N}^2}{1+\de_m}
= \frac{1}{H^2} (1+\de_m) \nabla^2 \Psi(\delta_{m\,NN}, \delta_{m\,N}, \delta_{m})\,.
\ee
{
We close this appendix by investigating the robustness of the top-hat assumption. 
The field profile can be computed by solving Eq.~\eqref{eq:x20}, which yields
\begin{align}
x = \frac{-B + \sqrt{B^2 - 4a C}}{2a},,
\label{xsol}
\end{align}
where the positive branch is chosen to ensure that the perturbed field $\pi$ does not diverge as $r \to \infty$, and
\be
B \equiv \bt_1 + \frac{\bt_2}{3 r^2} \frac{d}{dr}(r^3 A),,
\quad
C \equiv \mu A + \nu \dot{A},.
\ee
For the exterior region, $r > R$, where $R$ is the radius of the collapsing region, 
the mass defined in Eq.~\eqref{def:mass} is the total mass of the collapsing region and is therefore constant. 
Consequently, $A \propto 1/r^3$. 
Hence, $d(r^3 A)/dr = 0$ and $\dot{A} = -2\ddot\phi A/\dot\phi$. 
Let us define $c_1 \equiv (\mu - 2\ddot\phi\nu/\dot\phi) A$, 
so that Eq.~\eqref{xsol} can be integrated as
\be
\pi = \int_r^\infty r' x d r' 
=
\frac{R_*^2 \chi^2}{4 \al} 
\(-\bt_1 + \sqrt{\bt_1^2 - \frac{4 \al c_1}{\chi^3}} \frac{{}_2 F_1[-2/3, -1/2, 1/3, 4 \al c_1 / (\bt_1^2 \chi^3)]}{\sqrt{1 - 4 \al c_1 /(\bt_1^2 \chi^3)}}\right)\,,
\label{phiExt}
\ee
where ${}_2 F_1$ is the hypergeometric function and $\chi = r/r_*$. 
Outside the collapsing region, the Vainshtein radius associated with the total mass is denoted by $R_*$.

For the interior region, the matter is uniformly distributed according to the top-hat assumption, 
so that Eq.~\eqref{a-tophat} gives $d(r^3 A)/dr = 3A$, and
\be
\dot{A}
= \(- 2 \frac{\ddot\phi}{\dot\phi} + \frac{\dot\delta\rho_m}{\delta\rho_m}\)A
\equiv c_2 A,,
\ee
where, for simplicity, we assume that the leading contribution to $\dot\delta\rho_m/\delta\rho_m$ comes from background quantities. 
In this case, Eq.~\eqref{xsol} can be integrated as
\be
\pi =  \int_0^r r' x d r' 
= \frac{r_*^2\chi^2}{4 \al} \(-\bt_1 + \sqrt{\(\bt_1 + \frac{3 \bt_2}{\chi_R^3}\)^2 - \frac{4 \al c_3}{\chi_R^3}} - \frac{3 \bt_2}{\chi_R^3}\)
+ \pi_0\,,
\label{phiInt}
\ee
  where $c_3 \equiv \mu + c_2$, $\chi_R \equiv R/R_*$, and $\pi_0$ is the field value at the center of the collapsing region. 
  The expression for $\pi_0$ is obtained by matching $\pi$ from Eq.~\eqref{phiExt} to that from Eq.~\eqref{phiInt} at $r = R$.
The ratio ${\cal R} \equiv \pi_0/\pi(R)$ in the two limits of $\chi_R$ is
\be
{\cal R} \to 
\left\{\begin{array}{cc}
\frac{2 c_1 + c_3}{2 c_1} & \mbox{for}\,\, \chi_R \gg 1\\ 
1  & \mbox{for}\,\, \chi_R \ll 1 
\end{array}\right.\,,
\ee
where $\pi(R)$ is the field value at the surface, $r = R$, computed from Eq.~\eqref{phiExt}. 
It follows from the above equation that the fifth force inside the collapsing region is screened when the radius of the region is smaller than the Vainshtein radius.
We note that, if $\dot{A} = 0$, which yields $c_1 = c_3$, the above ratio agrees with the result in \cite{Schmidt:09}. 
However, for the DHOST theory considered here, the derivative of the field obtained from Eq.~\eqref{xsol} in the exterior region cannot be matched to that in the interior region unless $A$ at the surface of the region is small, 
which corresponds to the unscreened case.
From Eq.~\eqref{phiInt}, the field in the interior region is proportional to $r^2$, 
so that $\nabla^2\phi$ is constant. 
Consequently, $\nabla^2\psi$ from Eq.~\eqref{pos0} is also constant inside the collapsing region. 
In analogy with the discussion in \cite{Schmidt:09}, these results indicate that the initial top-hat profile is preserved throughout the collapse in the DHOST theories considered here, 
because the fifth force is proportional to $r$ and depends on the mass enclosed within that radius, similar to the force due to Newtonian gravity.
However, as discussed in the main text, the collapse can begin at an early stage when the screening mechanism is ineffective and become screened only at a later stage. 
Therefore, the top-hat assumption may not be fully applicable throughout the entire collapse process. 
The field profiles are shown in Fig.~\ref{fig:fp}.
In the plot, in the unscreened case, the field profile mimics the gravitational potential in Newtonian gravity, 
so that the fifth force mediated by the field depends on the radius and enclosed matter in a manner similar to the Newtonian gravitational force.
}
{
Even though the field profile is smooth inside the collapsing region, 
it does not significantly affect the quasi-static approximation used to derive Eq.~\eqref{ddotdel0}. 
This is because this equation describes the evolution of linear matter perturbations inside the Hubble radius, but on length scales that are assumed to be much larger than the radius of the collapsing region.
}

\begin{thebibliography}{99}

\bibitem{Langlois:2015cwa}
D.~Langlois and K.~Noui,
JCAP {\bf 02}, (2016) 034,   
[arXiv:1510.06930 [gr-qc]].

\bibitem{DHOST2016:2} 
M.~Crisostomi, K.~Koyama, and G.~Tasinato,
JCAP {\bf 04}, (2016) 044,    
[arXiv:1602.03119 [hep-th]].

\bibitem{DHOST2016:3} 
J.~Ben Achour, D.~Langlois, and K.~Noui,
Phys. Rev. D {\bf 93}, (2016) 124005,  
[arXiv: 1602.08398 [gr-qc]].

\bibitem{DHOST2016:4} 
J.~B.~Achour, M.~Crisostomi, K.~Koyama, D.~Langlois, and K.~Noui,
JHEP {\bf 12}, (2016) 100,     
[arXiv:1608.08135 [hep-th]].

\bibitem{DHOST2016:5} 
M.~Crisostomi, M.~Hull, K.~Koyama, and G.~Tasinato,
JCAP {\bf 03}, (2016) 038,     
[arXiv:1601.04658 [hep-th]].

\bibitem{deRham2016}
C.~de Rham and A.~Matas,
JCAP {\bf 06}, (2016) 041, 
[arXiv:1604.08638 [hep-th]].    

\bibitem{Monitor:2017mdv} 
B.~P.~Abbott {\it et al.},
Astrophys. J. Lett. {\bf 848}, (2017) L13,
[arXiv:1710.05834 [astro-ph.HE]].

\bibitem{TheLIGOScientific:2017qsa} 
B.~P.~Abbott {\it et al.},
Phys. Rev. Lett. {\bf 119}, (2017) 161101, 
[arXiv:1710.05832 [gr-qc]].


\bibitem{Creminelli2018} 
P.~Creminelli, M.~Lewandowski, G.~Tambalo, and F.~Vernizzi,
JCAP {\bf 12}, (2018) 025, 
[arXiv:1809.03484 [astro-ph.CO]]. 

\bibitem{Hirano:1903}
S.~Hirano, T.~Kobayashi and D.~Yamauchi,
Phys. Rev. D {\bf 99}, 104073 (2019),
[arXiv:1903.08399].

\bibitem{Hiramatsu:2020fcd}
T.~Hiramatsu and D.~Yamauchi,
Phys. Rev. D \textbf{102}, no.8, 083525 (2020)
doi:10.1103/PhysRevD.102.083525
[arXiv:2004.09520 [astro-ph.CO]].

\bibitem{Ma:1995ey}
C.~P.~Ma and E.~Bertschinger,
Astrophys. J. \textbf{455} (1995), 7-25
doi:10.1086/176550
[arXiv:astro-ph/9506072 [astro-ph]].

\bibitem{Kimura:2011dc} 
R.~Kimura, T.~Kobayashi and K.~Yamamoto,
Phys.\ Rev.\ D {\bf 85}, 024023 (2012)
[arXiv:1111.6749 [astro-ph.CO]].

\bibitem{Koyama:2013paa} 
K.~Koyama, G.~Niz and G.~Tasinato,
Phys.\ Rev.\ D {\bf 88}, 021502 (2013)
[arXiv:1305.0279 [hep-th]].

\bibitem{Crisostomi:1711}
M.~Crisostomi and K.~Koyama,
Phys. Rev. D {\bf 97}, 021301 (2018),
[arXiv:1711.06661 [astro-ph.CO]].

\bibitem{Schmidt:09}
Fabian Schmidt, Wayne Hu, and Marcos Lima,
Phys. Rev. D {\bf 81}, 063005 (2010),
[arXiv:0911.5178 [astro-ph.CO]].

\bibitem{Press:1974}
W.~H.~Press and P.~Schechter,
Astrophys.\ J.\ {\bf 187}, 425 (1974).

\bibitem{Viana:1999}
P.~T.~P.~Viana and A.~R.~Liddle,
Mon.\ Not.\ Roy.\ Astron.\ Soc.\ {\bf 303}, 535 (1999)
[astro-ph/9808244].

\bibitem{Sheth:1999}
R.~K.~Sheth and G.~Tormen,
Mon.\ Not.\ Roy.\ Astron.\ Soc.\ {\bf 308}, 119 (1999)
[astro-ph/9901122].

\bibitem{Reed:2006rw}
D.~Reed, R.~Bower, C.~Frenk, A.~Jenkins and T.~Theuns,
Mon.\ Not.\ Roy.\ Astron.\ Soc.\ {\bf 374}, 2 (2007)
[astro-ph/0607150].

\bibitem{Bulbul:2024}
E.~Bulbul \textit{et al.},
Astron.\ Astrophys.\ {\bf 685}, A106 (2024)
[arXiv:2402.08452 [astro-ph.CO]].
\end{thebibliography}
\end{document}